\PassOptionsToPackage{dvipsnames}{xcolor}
\documentclass[epjc3,twocolumn,nopacs]{svjour3}
\pdfoutput=1
\usepackage{amsmath,amsfonts,slashed,cite,orcidlink,listings,upquote}
\usepackage[switch]{lineno}
\usepackage[utf8]{inputenc}
\hypersetup{colorlinks=true,linkcolor=ForestGreen,citecolor=ForestGreen,filecolor=ForestGreen,urlcolor=ForestGreen}

\journalname{Eur. Phys. J. C}
\newcommand{\achilles}{{\sc Achilles}}

\newcommand{\comix}{{\sc Comix}}
\newcommand{\contur}{{\sc Contur}}
\newcommand{\feynrules}{{\sc FeynRules}}
\newcommand{\gosam}{{\sc GoSam}}
\newcommand{\herwig}{{\sc Herwig~7}}
\newcommand{\lanhep}{{\sc LanHep}}
\newcommand{\madanalysis}{{\sc MadAnalysis~5}}
\newcommand{\mgamc}{{\sc Mad\-Graph5\-\_aMC@NLO}}
\newcommand{\maddm}{{\sc Mad\-DM}}
\newcommand{\python}{{\sc Python}}
\newcommand{\recola}{{\sc Re\-co\-la~2}}
\newcommand{\sarah}{{\sc Sarah}}
\newcommand{\sherpa}{{\sc Sher\-pa}}
\newcommand{\whizard}{{\sc Whizard}}

\newcommand\identity{1\kern-0.25em\text{l}}
\newcommand{\ol}{\overline}
\newcommand{\bea}{\begin{eqnarray}}
\newcommand{\eea}{\end{eqnarray}}
\def\be{\begin{equation}}
\def\ee{\end{equation}}
\def\bsp#1\esp{\begin{split}#1\end{split}}

\def\ie{{\it i.e.}}
\def\etc{{\it etc.}}
\def\eg{{\it e.g.}}
\def\cf{{\it cf.}}

\definecolor{darkmagenta}{rgb}{0.55, 0.0, 0.55}
\definecolor{olivier}{rgb}{0.15, 0.15, 0.95}

\begin{document}

\title{UFO~2.0 - The `Universal Feynman Output' format}

\date{Received: date / Accepted: date}

\author{
  Luc Darmé\thanksref{addr5}%{e5,addr5}
  \and\
  C\'eline Degrande\thanksref{addr4}%{e9,addr4}
  \and\
  Claude Duhr\thanksref{addr1}%%{e1,addr1}
    \orcidlink{0000-0001-5820-3570}
  \and\
  Benjamin~Fuks\thanksref{e2,addr2}\orcidlink{0000-0002-0041-0566}
  \and\
  Mark~Goodsell\thanksref{addr2}%{e10,addr2}
    \orcidlink{0000-0002-6000-9467}
  \and\
  Gudrun Heinrich\thanksref{addr13}%{e17,addr13}
    \orcidlink{0000-0002-0834-3011}
  \and\
  Valentin Hirschi\thanksref{addr3}%{e3,addr3}
    \orcidlink{0000-0002-8908-6300}
  \and\
  Stefan~H{\"o}che\thanksref{addr10}%{e16,addr10}
    \orcidlink{0000-0002-1370-6059}
  \and\
  Marius~H\"ofer\thanksref{addr13}%{e18,addr13}
    \orcidlink{0000-0003-0009-9410}
  \and\
  Joshua~Isaacson\thanksref{addr10}%{e13,addr10}
    \orcidlink{0000-0001-6164-1707}
  \and\
  Olivier~Mattelaer\thanksref{addr4}%{e4,addr4}
  \and\
  Thorsten~Ohl\thanksref{addr8}%{e11,addr8}
    \orcidlink{0000-0002-7526-2975}
  \and\
  Davide Pagani\thanksref{addr6}%{e7,addr6}
    \orcidlink{0000-0002-0553-1105}
  \and\
  J\"urgen~Reuter\thanksref{addr9}%{e12,addr9}
    \orcidlink{0000-0003-1866-0157}
  \and\
  Peter~Richardson\thanksref{addr14}%{e19,addr14}
  \and\
  Steffen~Schumann\thanksref{addr12}%{e15,addr12}
    \orcidlink{0000-0003-0330-3990}
  \and\
  Hua-Sheng~Shao\thanksref{addr2}%{e6,addr2}
    \orcidlink{0000-0002-4158-0668}
  \and\
  Frank~Siegert\thanksref{addr11}%{e14,addr11}
    \orcidlink{0000-0002-2893-6412}
  \and\
  Marco~Zaro\thanksref{addr7}%{e8,addr7}
    \orcidlink{0000-0002-3279-7355}
}

%\thankstext{e5}{E-mail: {\color{ForestGreen}l.darme@ip2i.in2p3.fr}}
%\thankstext{e9}{E-mail: {\color{ForestGreen}celine.degrande@uclouvain.be}}
%\thankstext{e1}{E-mail: {\color{ForestGreen}cduhr@uni-bonn.de}}
\thankstext{e2}{E-mail: {\color{ForestGreen}fuks@lpthe.jussieu.fr}}
%\thankstext{e10}{E-mail: {\color{ForestGreen}goodsell@lpthe.jussieu.fr}}
%\thankstext{e17}{E-mail: {\color{ForestGreen}gudrun.heinrich@kit.edu}}
%\thankstext{e3}{E-mail: {\color{ForestGreen}valentin.hirschi@gmail.com}}
%\thankstext{e16}{E-mail: {\color{ForestGreen}shoeche@fnal.gov}}
%\thankstext{e18}{E-mail: {\color{ForestGreen}marius.hoefer@kit.edu}}
%\thankstext{e13}{E-mail: {\color{ForestGreen}isaacson@fnal.gov}}
%\thankstext{e4}{E-mail: {\color{ForestGreen}olivier.mattelaer@uclouvain.be}}
%\thankstext{e11}{E-mail: {\color{ForestGreen}ohl@physik.uni-wuerzburg.de}}
%\thankstext{e7}{E-mail: {\color{ForestGreen}davide.pagani@bo.infn.it}}
%\thankstext{e12}{E-mail: {\color{ForestGreen}juergen.reuter@desy.de}}
%\thankstext{e19}{E-mail: {\color{ForestGreen}peter.richardson@durham.ac.uk}}
%\thankstext{e15}{E-mail: {\color{ForestGreen}steffen.schumann@phys.uni-goettingen.de}}
%\thankstext{e6}{E-mail: {\color{ForestGreen}huasheng.shao@lpthe.jussieu.fr}}
%\thankstext{e14}{E-mail: {\color{ForestGreen}frank.siegert@cern.ch}}
%\thankstext{e8}{E-mail: {\color{ForestGreen}marco.zaro@mi.infn.it}}

\institute{
  Institut de Physique des 2 Infinis de Lyon (IP2I),
UMR5822, CNRS/IN2P3, F-69622 Villeurbanne Cedex, France \label{addr5}
  \and\
  Universit\'e catholique de Louvain,
  Center for particle physics and phenomenology (CP3),
  Chemin du cyclotron, 2, B-1348 Louvain-La-Neuve, Belgium\label{addr4}
  \and\
  Bethe Center for Theoretical Physics, Universit\"at Bonn, D-53115, Germany \label{addr1}
  \and\
  Laboratoire de Physique Th\'eorique et Hautes Energies (LPTHE), UMR 7589, Sorbonne Universit\'e et CNRS, 4 place Jussieu, 75252 Paris Cedex 05, France\label{addr2}
  \and\
  Karlsruhe Institute of Technology, Institute for Theoretical Physics, Wolfgang-Gaede-Str. 1, 76131 Karlsruhe, Germany\label{addr13}
  \and\
  Institute for Theoretical Physics, University of Bern, Sidlerstrasse 5, 3012 Bern, Switzerland\label{addr3}
  \and\
  Fermi National Accelerator Laboratory, Batavia, IL 60510, USA\label{addr10}
  \and\
  University of Würzburg, Institut für Theoretische Physik
  und Astrophysik, Emil-Hilb-Weg 22, 97074 W\"urzburg, Germany\label{addr8}
  \and\
  INFN, Sezione di Bologna, Via Irnerio 46, 40126 Bologna, Italy\label{addr6}
  \and\
  Deutsches Elektronen-Synchrotron DESY, Notkestr.\ 85, 22607 Hamburg, Germany\label{addr9}
  \and\
  Institute for Particle Physics Phenomenology, Durham University, Durham, UK\label{addr14}
  \and\
  Institute for Theoretical Physics, 
  Georg-August-University G{\"o}ttingen,
  Friedrich-Hund-Platz 1, 
  37077 G{\"o}ttingen, 
  Germany\label{addr12}
  \and\
  Institut f\"ur Kern- und Teilchenphysik, TU Dresden, 01069 Dresden, Germany\label{addr11}
  \and\
  Universit\`a degli Studi di Milano \& INFN, Sezione di Milano, Via Celoria 16, 20133 Milano, Italy \label{addr7}
}

\maketitle

\begin{abstract}
We present an update of the \textit{Universal FeynRules Output} model format, commonly known as the UFO format, that is used by several automated matrix-element generators and high-energy physics software. We detail different features that have been proposed as extensions of the initial format during the last ten years, and collect them in the current second version of the model format that we coin the {\it Universal Feynman Output} format. Following the initial philosophy of the UFO, they consist of flexible and modular additions to address particle decays, custom propagators, form factors, the renormalisation group running of parameters and masses, and higher-order quantum corrections.
\end{abstract}

\vspace*{-18.5cm}
  \noindent {\small\texttt{BONN-TH-2023-03, DESY-23-051, FERMILAB-PUB-23-138-T, KA-TP-06-2023, MCNET-23-06, P3H-23-023, TIF-UNIMI-2023-11}}
\vspace*{17.5cm}

\section{Introduction}
During the last 25--30 years, several high-energy physics software packages have been developed to explore the electroweak scale and get information on the possible physics beyond the Standard Model (BSM). Typical examples of such programs target the simulation of events at high-energy collider, fixed-target or neutrino experiments, total and differential cross section calculations for many processes in the Standard Model (SM) and beyond it, as well as the computation of dark matter observables. These software tools generally require as input, in one form or another, the particle spectrum of the model, the list and the values of all parameters that appear in its Lagrangian, as well as the list of all interaction vertices among the different particles. Historically, each program followed its own format to input the model information, with its own conventions and restrictions on the supported structures in a Lagrangian. This severely limited the portability of a model, and consequently multiplied the workload for the implementation and validation into several tools as advocated in \cite{Christensen:2009jx}.

The UFO format~\cite{Degrande:2011ua} was proposed as a solution to this issue, by introducing a new way to pass model information to high-energy physics software. Its goal is to provide a flexible and fully generic format that goes beyond existing formats in the sense that no assumption on the supported structures appearing in the model is enforced. {\it All} the model information is stored in an abstract form, \ie\ independent of the  software. It is then up to the tool using the UFO model to enforce their restrictions at run time. The UFO representation of the particle physics model has been chosen to rely on \python\ objects defined through a set of attributes that encode physical properties so that the model could be straightforwardly accessed and parsed by any high-energy physics tool. One of the advantages of the design choices made is that the UFO is modular. Additionally the format is easily expandable to include new pieces of information not originally considered. These design choices allowed later developments that permitted the inclusion of decay width information~\cite{Alwall:2014bza}, modifications of the propagators associated with any given field~\cite{Christensen:2013aua}, renormalisation group running effects impacting some of the model's parameters and masses~\cite{Aoude:2022aro}, and ingredients relevant to higher-order perturbative calculations in quantum field theories~\cite{Degrande:2014vpa}.

With the present paper, we take the opportunity to collect all these recent developments in a single document, describe (for the first time) how to embed form factors in UFO models, and how to include missing information relevant to the automated calculations of electroweak corrections in the Sudakov approximation for collider processes. In section~\ref{sec:ufo_format}, we begin with a general description of the UFO format. We provide additional details on the philosophy of the UFO format, describe the structure of how the model information is organised in several \python\ files, and put a particular emphasis on (optional) recommendations useful for making UFO models traceable. Section~\ref{sec:ufo_mandatory} is dedicated to the original UFO format, and we describe all mandatory files that should be included in a UFO model. In section~\ref{sec:ufo_optional}, we detail how optional components can be added to a UFO model, and describe all existing options. Finally, section~\ref{sec:ufo_nlo} focuses on higher-order computations and how ingredients relevant to this context could be included in the UFO format, both in general and for the specific case of electroweak  corrections in the Sudakov approximation. We summarise our work in section~\ref{sec:conclusions}.

\section{The UFO format}
\label{sec:ufo_format}
\subsection{The evolution of the UFO format}

The aim of this section is to provide a general overview of the UFO~2.0 format for new physics models, that we propose to call the \emph{Universal Feynman Output} (UFO) format in order to distinguish it from the initial version~\cite{Degrande:2011ua} released a decade ago.  In the following, we emphasise the philosophy behind the UFO format, as well as its general structure. The content of the different files included in a UFO model and the associated syntax are discussed in more detail in dedicated subsequent sections. 

A UFO model consists of a set of \python\ files that can be used with a large class of publicly available computer packages relevant for high-energy physics calculations. The UFO format has been built around the philosophy that a model implementation should be independent of the software tool that uses it. This makes it possible to have a single model implementation working across different computer codes and platforms, making it relevant for assessing the phenomenology relevant for different classes of experiments (targeting, for instance, dark matter, high-energy collider or neutrino experiments). The UFO standard achieves this by representing the model information, namely the model's particles, parameters and vertices, in terms of \python\ objects whose attributes collect their properties. It is then up to the computer code that uses the model implementation to read in these files, and to process their content correctly. In case the code has restrictions on the type of models, an exception is raised and informs the user that the implementation cannot be reliably used.

The first version of the UFO format~\cite{Degrande:2011ua} was released a decade ago. It has changed the way particle physics models in general, and theories beyond the SM in particular, are implemented in high-energy physics software. Whilst the UFO format initially targeted specifically the implementation of particle physics models in matrix element and event generators dedicated to studies at the leading-order (LO) accuracy in perturbation theory, it is currently supported by a larger list of high-energy physics software tools. This list includes \achilles~\cite{Isaacson:2021xty,Isaacson:2022cwh}, \comix~\cite{Gleisberg:2008fv}, \contur~\cite{Butterworth:2016sqg}, \gosam~\cite{Cullen:2011ac, Cullen:2014yla}, \herwig~\cite{Bahr:2008pv,Bellm:2015jjp}, \madanalysis~\cite{Conte:2012fm,Conte:2018vmg}, \maddm~\cite{Ambrogi:2018jqj,Arina:2020kko,Arina:2021gfn}, \mgamc~\cite{Alwall:2014hca,Frederix:2018nkq}, \recola~\cite{Denner:2017bdv}, \sherpa~\cite{Hoche:2014kca, Sherpa:2019gpd} and \whizard~\cite{Moretti:2001zz, Kilian:2007gr, Christensen:2010wz}.

Since its inception, the UFO format underwent several extensions to accommodate the specification of additional model information which is not part of its original documentation~\cite{Degrande:2011ua}, such as those mentioned above. In particular, the documentation related to automated next-to-leading order (NLO) computations has never been collected in a single document, despite being at the heart of the frameworks introduced in \cite{Alwall:2014hca, Degrande:2014vpa, Frederix:2018nkq}. UFO models suitable for NLO calculations have in addition become standard in high-energy phenomenology during the last decade, a large variety of NLO-compatible UFO models being now available (especially from the {\sc FeynRules} model database\footnote{See the webpage \url{http://feynrules.irmp.ucl.ac.be/wiki/NLOModels}.}). Although all extensions mentioned above are already being used by several codes, there is no official documentation of the structure of the UFO format beyond the original proposal. The main purpose of this document is therefore to provide an update of the UFO documentation, which contains all the features relevant for computations beyond LO accuracy.

Before starting to discuss the general outline of a UFO model implementation, let us first make a comment about the name. Originally, the acronym UFO stood for `Universal FeynRules Output'. The origin of this name can be traced back to the fact that in its original conception UFO files were produced by \feynrules~\cite{Alloul:2013bka} only. For a few years now, UFO files can also be generated from a user-defined Lagrangian by other computer codes such as \lanhep~\cite{Semenov:2008jy,Semenov:2014rea} and \sarah~\cite{Staub:2013tta,Goodsell:2017pdq}. For this reason we deem it more appropriate to remove the explicit reference to \feynrules\ from the name of the UFO format, and the acronym UFO henceforth now stands for `Universal Feynman Output'.

\subsection{General file structure of the UFO}\label{sec:generalstructure}
In the remainder of this paper we discuss in detail the structure of the files contained in a UFO model. All the files, collected in a single directory, must be valid \python\ files, therefore with a file extension {\tt .py}. Whereas most files are model-specific and contain the definition of the objects relevant to each model (\eg\ particles and parameters), some of the files are model-independent and contain \python-code objects defining, for instance, the \python\ classes used in a UFO model. 

The following model-specific files are mandatory in any valid UFO model directory,
\begin{itemize}
    \item \verb+particles.py+
    \item \verb+parameters.py+
    \item \verb+vertices.py+
    \item \verb+lorentz.py+
    \item \verb+couplings.py+
    \item \verb+coupling_orders.py+
    \item \verb+function_library.py+
\end{itemize}
These files contain the basic definitions related to a model. If a model is to be used for computations beyond LO accuracy, three extra files are mandatory and must be included,
\begin{itemize}
    \item \verb+CT_vertices.py+
    \item \verb+CT_couplings.py+
    \item \verb+CT_parameters.py+
\end{itemize}
Moreover, in the specific case of electroweak corrections in the high-energy (Sudakov) approximation, this list must be complemented by an additional file that is described in this document for the first time,
\begin{itemize}
    \item \verb+CT_ewcasimirs.py+
\end{itemize}
Finally, every UFO directory may contain certain optional files, which specify additional model information,
\begin{itemize}
    \item \verb+form_factors.py+
    \item \verb+decays.py+
    \item \verb+propagators.py+
    \item \verb+running.py+
\end{itemize}
The content of these files is described in detail in section~\ref{sec:ufo_optional}, so that we limit ourselves here to highlighting some features that are common to all of them. 

Each file defines a list of objects. The classes that can be used are predefined and included in the mandatory file \verb+object_library.py+ (see below), and only standard \python\ syntax is allowed. Several of the files define analytic expressions for interaction vertices or coupling constants in the theory. All standard arithmetic operations in \python\ can be used to write such analytic expressions in the UFO format, augmented by some special symbols whose meaning is described in subsequent sections together with the precise syntax.

Besides these model-specific files which are at the heart of every UFO implementation, there are a couple of mandatory model-independent files that need to be included in every valid UFO directory,
\begin{itemize}
    \item \verb+object_library.py+
    \item \verb+__init__.py+
\end{itemize}
together with the optional file
\begin{itemize}
    \item \verb+write_param_card.py+
\end{itemize}
that has a specific practical use.

As already mentioned, the file \verb+object_library.py+ contains the definition of all classes used in a UFO model. It includes several lists providing easy access to the {\it full} content of the model within the code. In other words, {\it all} declared objects within a UFO must appear in these lists. The list \verb+all_particles+ collects all particle declarations (as instances of the \verb+Particle+ class; see section~\ref{sec:particles}), and the list \verb+all_parameters+ gathers all parameter declarations (as instances of the \verb+Parameter+ class; see section~\ref{sec:parameters}). The elements required for the description of the model interactions (see section~\ref{sec:interactions}) are spread over the list \verb+all_vertices+ that collects all vertex declarations (as instances of the \verb+Vertex+ class; see section~\ref{sec:interactions}), the list \verb+all_couplings+ that collects all coupling declarations (as instances of the \verb+Coupling+ class), the list \verb+all_lorentz+ that includes all Lorentz tensors appearing in the model vertices (declared as instances of the \verb+Lorentz+ class), and finally the additional list \verb+all_coupling_orders+ that contains a list of tags allowing certain vertices of the model to be flagged (these tags being declared as instances of the \verb+CouplingOrder+ class). In addition, the \python\ file \verb+object_library.py+ also includes a list \verb+all_functions+ whose role is to gather all \verb+Function+ objects instantiated in the model (see section~\ref{sec:fct}).

The content of the files \verb+function_library.py+ and \verb+write_param_card.py+ is detailed in sections~\ref{sec:ufo_mandatory} and \ref{sec:ufo_optional}. We only focus here on the file \verb+__init__.py+.
This file identifies the content of a UFO directory as a valid \python\ module that can be loaded with the standard command {\tt import}, and it may contain any valid \python\ command that should be evaluated when the model is loaded. In particular, the file \verb+__init__.py+ imports all other \python\ files relevant to the model, and it additionally allows users to add general information about the model, as shown in the following example:
\begin{verbatim}
    __author__         = "H. Solo, C. Bacca"
    __date__           = "06.03.2023"
    __model_version__  = "1.0"
    __arxiv__          = "2304.NNNNN"
    __UFO_version__    = "2.X"
    __python_version__ = [2,3]
\end{verbatim}
The first three variables (\verb+__author__+, \verb+__date__+ and \verb+__model_version__+) provide information on the implementation and its author(s), whereas the \verb+__arxiv__+ variable enables the connection of a given UFO model to a publication released on the arXiv. Setting the variable \verb+__UFO_version__+ to \verb+"2.X"+ indicates that the mo\-del implementation includes features documented in the present paper, and the \verb+__python_version__+ variable refers to the version of \python\ with which the UFO is compatible (namely \verb+2+ and/or \verb+3+ at present time). While such an electronic signature of the model is not mandatory, we recommend users to include it for traceability reasons. Depending on the moment at which a UFO model has been generated and that at which it is used within a code, incompatibilities between \python\ versions may occur. While we suggest to update existing UFO models so that they become \python~3 compatible, it is up to the code using UFOs to make sure that \python\ version compatibility is addressed properly and internally. For instance, \gosam, \herwig\ and \mgamc\ convert UFO models compatible with \python~2 to their \python~3 equivalent in order to use them.

Finally, users can include information on the gauges available for the model implementation. This is achieved through the variable {\tt gauge} that contains a list of integers, as for instance in 
\begin{verbatim}
    gauge = [0,1]
\end{verbatim}
The value \verb+0+ refers to the unitarity gauge, whereas the value \verb+1+ stands for the Feynman gauge. Other integer values are allowed, provided that they are consistently defined in the UFO model, in particular through appropriate definitions in the files \verb+parameters.py+ (for gauge parameters like $\xi$ in the $R_\xi$ gauge) and  \verb+propagators.py+ (for custom propagator expressions).

\section{Mandatory components}
\label{sec:ufo_mandatory}
The dynamics of a particle physics model at tree-level is encoded in the UFO format within a small set of mandatory files. This contains the description of the particle spectrum (\verb+particles.py+), the model parameters (\verb+parameters.py+) and the different interactions between the model particles (whose implementation is spread over the three files \verb+vertices.py+, \verb+couplings.py+ and \verb+lorentz.py+). In addition, two extra files are necessary. The first of them, \verb+coupling_orders.py+, details tags allowing certain vertices of the model to be flagged, whereas  the last one, \verb+functions_library.py+, is dedicated to the implementation of user-defined functions that can be used anywhere in the UFO model. The content of all these files is described in the following subsections.

\subsection{Particles}\label{sec:particles}

All physical particles of a model are declared as instances of the class \verb+Particle+ in the file \verb+particles.py+. UFO models are generally defined in terms of the physical, propagating, mass eigenstates. Unphysical gauge eigenstates and non-propagating auxiliary fields are thus ignored in most implementations, with the exception of optional ghost and Goldstone fields that may be needed depending on the gauge chosen. However, it is always possible to include specific auxiliary fields in an implementation if needed (see also the end of this subsection). A \verb+Particle+ object is defined through various attributes specifying the particle name and properties, including its quantum numbers. As an illustration, we consider a possible UFO implementation for a heavy top quark $t'$, 
\begin{lstlisting}
tp = Particle(
   pdg_code     = 8,
   name         = 'tp',
   antiname     = 'tp$\sim$',
   spin         = 2,
   color        = 3,
   mass         = Param.MTP,
   width        = Param.WTP,
   texname      = 'tp',
   antitexname  = 'tp$\sim$',
   charge       = 2/3,
   LeptonNumber = 0
)\end{lstlisting}
The particle is identified by its name (the \verb+name+ attribute taken to be \verb+tp+ in the present example), its spin and colour representations (given as the value of the \verb+spin+ and \verb+color+ attributes), its mass and width (the value of the \verb+mass+ and \verb+width+ attributes, given in GeV) and its electric charge (given as the value of the \verb+charge+ attribute, in units of the proton's electric charge). The \verb+tp+ symbol that appears on the left-hand side of the equality represents a unique \python\ identifier that is further used internally within the model to refer to that particle. It has thus to follow \python\ requirements for names of variables. However, this identifier will {\it not} appear within any of the lists introduced in the file \verb+object_library.py+, that include instead the objects themselves. All \verb+Particle+ objects instantiated within a UFO model must therefore have unique \verb+name+ attribute values, as this is how they should be referred to within any code using UFOs, in addition to unique \python\ identifiers. In addition, such a constraint holds for all the other classes of objects introduced below: two instances of a given class must have different \verb+name+ attributes\footnote{In the case where a given UFO model has to be used within a toolchain involving a parton showering and hadronisation program, it is best to also avoid using the names of standard mesons and hadrons, like \texttt{eta} and \texttt{sigma}.}.

In the UFO conventions, the spin representation has to be provided in the $2s+1$ form where $s$ denotes the particle spin. Whereas any $s$ values are allowed at the UFO level, none of the tools currently employing UFO models are compliant with spins $s>2$. Moreover, setting \verb+spin = -1+ identifies ghost fields. Similarly, whereas users are free to assign any colour representation for a particle in a model, tools currently making use of UFO models support at most the trivial, (anti)fundamental, (anti)sextet and adjoint representations. These choices can be made by setting the \verb+color+ attribute to 1, $\pm3$, $\pm6$ and 8. 

Information on the particle mass and width are provided by referring to the corresponding model parameters. In the considered example, the \verb+mass+ and \verb+width+ attributes of the \verb+tp+ particle are set to \verb+MTP+ and \verb+WTP+, that are both declared in the file \verb+parameters.py+ (see section~\ref{sec:parameters}). Parameter declarations must consequently be imported prior to the declaration of any particle, \ie\ by inserting at the beginning of the file \verb+particles.py+:\footnote{For models that are \python~3-compatible, this should read, according to standard conventions:\\ \texttt{from . import parameters as Param}}
\begin{verbatim}import parameters as Param
\end{verbatim}

The UFO conventions allow users to associate a particle with its corresponding antiparticle. This is achieved through the \verb+antiname+ attribute of the \verb+Particle+ class, which must be set to the name of the \verb+Particle+ object representing the antiparticle. The latter is itself declared either as above (with some of the attribute values swapped or modified) or through the more economical method \verb+anti()+,
\begin{verbatim}tp__tilde__ = tp.anti()\end{verbatim}
This method of the \verb+Particle+ class is defined in the file \verb+object_library.py+, and it automatically instantiates an antiparticle from the corresponding particle object. The \TeX\ version of the particle and antiparticle names are respectively provided as the value of the \verb+texname+ and \verb+antitexname+ attributes. In the case of a self-conjugate particle, all antiparticle attributes must be set to the same value as their particle counterparts. 

Most high-energy physics programs dealing with particles often internally identify them through their Particle Data Group (PDG) identifiers~\cite{ParticleDataGroup:2022pth}. In the UFO format, such an identifier is stored as the value of the \verb+pdg_code+ attribute of the \verb+Particle+ class, that has been chosen to be $8$ in the $t'$ example considered. While users can technically assign any code to any particle, many programs employing UFO models have the standard identifiers provided in the PDG review~\cite{ParticleDataGroup:2022pth} hard-coded for common BSM particles. Inconsistent choices may therefore lead to unexpected behaviours of these tools. We recommend users to make use of existing identifiers for particles already listed in the PDG review, and new non-used identifiers otherwise.

While all the attributes described above are mandatory, additional optional attributes (like the attribute \verb+LeptonNumber+ in the \verb+tp+ example considered) can be included. The UFO format includes the five predefined attributes \verb+line+, \verb+goldstone+, \verb+propagating+, \verb+counterterm+ and \verb+propagator+. The first three attributes indicate how to draw the particle propagator in a Feynman diagram (the possible self-explanatory values of the attribute \verb+line+ being \verb+'dashed'+, \verb+'dotted'+, \verb+'straight'+, \verb+'wavy'+, \verb+'curly'+, \verb+'scurly'+, \verb+'swavy'+ and \verb+'double'+), whether the particle is a Goldstone boson (\verb+'true'+) or not (default, \verb+'false'+), and whether it consists of a physical field that propagates (\verb+'true'+, default) or of a non-propagating auxiliary field (\verb+'false'+). Information of the last two of these predefined optional attributes, \verb+counterterm+ and \verb+propagator+, is provided in sections~\ref{sec:ufo_nlo} and \ref{sec:propagators} respectively.

Finally, any extra attribute appearing in the instantiation of a \verb+Particle+ object (like $U(1)$ quantum numbers such as \verb+LeptonNumber+ in the above example) represents a model-dependent quantum number whose sign changes under the action of the \verb+anti()+ method relevant for antiparticle objects.

%:
\subsection{Parameters}\label{sec:parameters}
Model parameters (masses, couplings, mixing matrix elements, \etc) are declared as instances of the \verb+Parameter+ class in the file \verb+parameters.py+. The UFO syntax distinguishes external and internal parameters. The former are the free parameters for which numerical values have to be provided by the user, while the latter are derived quantities related to other parameters (internal and/or external) via algebraic relations. Accordingly, a numerical value has to be provided for an external parameter whilst an analytical formula has to be given for an internal parameter. The UFO format also includes a third class of parameters, called constant parameters, that are similar to external parameters except that their value cannot be changed by the user. Equivalently, such constant parameters could also be declared as internal parameters for which the analytical expression is equal to a numerical value. Consequently, the only possibility to modify the value of a constant parameter is to edit directly the file \verb+parameters.py+.

A typical declaration of an external parameter would be
\begin{verbatim}
tb = Parameter(
   name     = 'tb',
   nature   = 'external',
   type     = 'real',
   value    = 10.,
   texname  = '\\text{tb}',
   lhablock = 'HMIX',
   lhacode  = [ 2 ]
)\end{verbatim}
In this example, we considered the parameter $\tan\beta$ that is defined as the ratio of the vacuum expectation values of the neutral Higgs fields in two-Higgs-doublet models, and that is often taken as one of the external parameters describing the Higgs sector of the model.

The above expression declares an instance of the \verb+Parameter+ class called \verb+tb+ (the value of the \verb+name+ attribute being \verb+tb+). The nature of this parameter is external, as indicated by the value of the \verb+nature+ attribute (that has been set to \verb+external+). In contrast, this attribute has to be fixed to \verb+internal+ or \verb+constant+ for internal and constant parameters respectively (see below for dedicated examples). In the above instantiation, the \verb+tb+ parameter is imposed to be real, since external parameters {\it must} all be real numbers. This is achieved through the attribute \verb+type+ whose value is set to \verb+real+ (the other possible option being \verb+complex+). Consequently, the value of the \verb+value+ attribute is a floating-point number ($10$ in the above example). In addition, the \TeX\ version of the parameter name must be specified, as for particle names (see section~\ref{sec:particles}), by setting accordingly the \verb+texname+ attribute. In principle users can choose the name of the parameters of a model freely, some parameter names are reserved as lying at the heart of higher-order calculations. We refer to section~\ref{sec:ufo_nlo} for more information.

The last two attributes in the above declaration, namely \verb+lhablock+ and \verb+lhacode+, refer to the way in which external parameters are organised, following conventions generalising the Supersymmetry Les Houches Accord (SLHA) format~\cite{Skands:2003cj,Allanach:2008qq}. In this scheme, the numerical values of all the model parameters are collected into specific \emph{blocks}, and each parameter is identified inside a block by one or more integer numbers called \emph{counters}. These counters consist of a single integer for scalar parameters, and in a sequence of integers for tensor parameters, the integers corresponding to the tensor indices. Moreover, all the elements of a given tensor must be part of the same Les Houches block. In the case of the $\tan \beta$ declaration above, such a Les Houches structure would correspond to
\begin{verbatim}
Block HMIX
    2 1.000000e+01 # tb
\end{verbatim}
In the UFO conventions, the name of the block (\verb+'HMIX'+) is passed as the value of the \verb+lhablock+ attribute, while the counter (\verb+[ 2 ]+) is given as an array through the value of the \verb+lhacode+ attribute. In the SLHA-based format, the numerical value of the parameter (\verb"1.00e+01" here) is given after the counter, followed by an optional comment (referring in the above example to the parameter name).

Whereas the user can freely choose the names of the various Les Houches blocks and how the counters are organised, the SM parameters have to be correctly identified by any tool using a UFO model. For instance, if the SM input parameters include the inverse of the electromagnetic coupling constant at the $Z$-pole $\alpha^{-1}(m_Z)$, the Fermi constant $G_F$ and the strong coupling constant at the $Z$-pole $\alpha_s(m_Z)$, then they have to be defined as the first three entries of the \verb+SMINPUTS+ block, the electromagnetic and strong coupling constants $\alpha$, $e$ and $g_s$ being in this case internal quantities.\footnote{There is no restriction on the adopted electroweak scheme. Any choice has its conventions in terms of external and internal parameters, and on the manner to encode them in a Les Houches structure. On the other hand, if the model allows for the calculation of NLO electroweak corrections, then the corresponding renormalisation conditions have to be consistently implemented, as discussed in section~\ref{sec:ufo_nlo}.} In addition, masses and widths must be assigned to the blocks \verb+MASS+ and \verb+DECAY+, the counter being the PDG code of the particle. We refer to the \feynrules\ manual~\cite{Alloul:2013bka} and the description of the SLHA format~\cite{Skands:2003cj,Allanach:2008qq} for more information on these conventions. Finally, UFO models suitable for higher-order calculations should include the blocks {\tt LOOP} and {\tt TECHNICAL}, that contain specific parameters relevant for programs handling calculations beyond LO. Their role is detailed in section~\ref{sec:ufo_nlo}.

This SLHA-like structure associated with the organisation of the external parameters is irrelevant for internal and constant parameters, so that instantiation of the latter does not require users to provide values for the \verb+lhablock+ and \verb+lhacode+ attributes. Moreover, constant and internal parameters can be complex quantities, in contrast with external parameters. This is indicated by setting the \verb+type+ attribute to the value \verb+complex+. In the case of internal parameters, the attribute \verb+value+ is fixed to a valid algebraic \python\ expression represented by a string. This formula can depend on any external, constant or internal parameter already declared in the file \verb+parameters.py+ (\ie\ on any parameter appearing before in the file). For constant parameters, a numerical value has to be provided instead. 

As an illustrative example, we show how to define the cosine of the $\beta$ angle. It can be derived from $\tan\beta$ (defined as an external parameter earlier), and can be declared in a UFO model as
\begin{verbatim}
cbeta = Parameter(
   name    = 'cbeta',
   nature  = 'internal',
   type    = 'real',
   value   = 'math.cos(math.atan(tb))',
   texname = '\\cos\\beta'
)\end{verbatim}
after having properly imported the {\tt math} module.

\subsection{Interactions} \label{sec:interactions}
The cornerstone of the UFO format consists of the way in which interactions are implemented, following their decomposition in a colour $\otimes$ spin space. Any generic vertex ${\cal V}$ involving the interaction of $n$ external particles $\varphi_i^{\ell_i a_i}(p_i)$ ($i=1, \ldots, n$) with spin indices $\ell_i$ (equivalently denoting Dirac and Lorentz indices), colour indices $a_i$ and four-momenta $p_i$, could be decomposed as
\be\bsp
  &\quad {\cal V}^{a_1\ldots a_n, \ell_1\ldots\ell_n}(p_1,\ldots,p_n) =\\
    &\qquad\quad
    \sum_{i,j}C_i^{a_1\ldots a_n} \ G_{ij}\times  L_j^{\ell_1\ldots\ell_n}(p_1,\ldots,p_n) \,.
\esp\label{eq:decomp}\ee
In this expression, the vertex ${\cal V}$ is decomposed into a set of colour structures $C_i^{a_1\ldots a_n}$ and spin structures $L_j^{\ell_1\ldots\ell_n}(p_1,\ldots,p_n)$, that are given as tensors in colour and spin space respectively. After considering all the model interactions, the resulting ensemble of structures defines a colour and spin basis allowing for the decomposition of any of the model vertices. Eq.~\eqref{eq:decomp} hence underlines an economical way to define all the interactions of the model, since a given spin or colour tensor could be used in several vertices. The set of coordinates associated with a specific vertex in the colour $\otimes$ spin basis are given by the coupling strengths $G_{ij}$. In version~2.0, the UFO format only supports unbroken gauge groups that comprise a single copy of $SU(3)$ and any number of $U(1)$ factors, such as in the SM after electroweak symmetry breaking.

As an example, we consider the four-scalar interaction between right-handed up squarks and antisquarks of the Minimal Supersymmetric Standard Model, whose associated Feynman rule is given by:\\
\begin{tabular}{lcl}
  \parbox{0.2\columnwidth}{\includegraphics[width=.385\columnwidth]{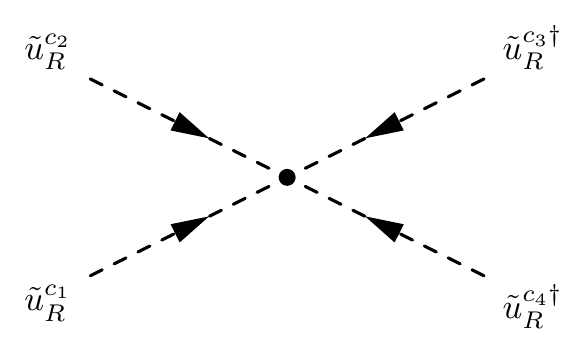}} &
  \parbox{0.13\columnwidth}{$~$} &
  \parbox{0.68\columnwidth}{$\displaystyle
    -\frac{4 i e^2}{9 c_W^2}
     \Big[\delta^{\bar c_4}{}_{c_1}\delta^{\bar c_3}{}_{c_2}\!+\!\delta^{\bar c_3}{}_{c_1}\delta^{\bar c_4}{}_{c_2}\Big]
     \\[.2cm]
    -i g_s^2\Big[(T^a)^{\bar c_3}{}_{c_2} (T^a)^{\bar c_4}{}_{c_1} \\ \hspace*{1.3cm}+ (T^a)^{\bar c_3}{}_{c_1} (T^a)^{\bar c_4}{}_{c_2}\Big]$,}
\end{tabular}
where $c_1$ and $c_2$ ($\bar c_3$ and $\bar c_4$) denote the fundamental (anti-fundamental) colour indices of the two squarks (antisquarks), $a$ is a summed adjoint colour index, and $c_W$ is the cosine of the electroweak mixing angle. Moreover, $T^a$ stands for the $SU(3)$ generators in the fundamental representation, and $g_s$ and $e$ are the strong and electromagnetic coupling constant respectively. The UFO decomposition of this vertex can be written as%
\renewcommand{\arraystretch}{1.25}%
\be\small{\bsp
  &\bigg(\delta^{\bar c_4}{}_{c_1}\delta^{\bar c_3}{}_{c_2}\quad
        \delta^{\bar c_3}{}_{c_1}\delta^{\bar c_4}{}_{c_2}\quad
        (T^a)^{\bar c_3}{}_{c_2} (T^a)^{\bar c_4}{}_{c_1}\quad
        (T^a)^{\bar c_3}{}_{c_1} (T^a)^{\bar c_4}{}_{c_2} \bigg)\\
   &\hspace{4cm}\times
   \begin{pmatrix} -(4 i e^2)/(9 c_W^2)\\-(4 i e^2)/(9 c_W^2)\\
     -ig_s^2\\-ig_s^2\end{pmatrix} \times
     \begin{pmatrix} 1\end{pmatrix}\,.
\esp}\label{eq:UFOex}\ee
\renewcommand{\arraystretch}{1.0}
The colour basis ${\cal C} = \Big(C_i^{c_1c_2\bar c_3 \bar c_4}\Big)$ contains four elements,
\be\bsp
  &{\cal C} = \bigg( 
    \delta^{\bar c_4}{}_{c_1}\delta^{\bar c_3}{}_{c_2},\
    \delta^{\bar c_3}{}_{c_1}\delta^{\bar c_4}{}_{c_2},\\ & \hspace{2cm}
    (T^a)^{\bar c_3}{}_{c_2} (T^a)^{\bar c_4}{}_{c_1},\ 
    (T^a)^{\bar c_3}{}_{c_1} (T^a)^{\bar c_4}{}_{c_2}
  \bigg) \,,
\esp\label{eq:colobasis}\ee
whereas the spin basis ${\cal L}$ contains a single element
\be
  {\cal L} = \big(1 \big)\,.
\ee 
Here the coordinates ${\cal G}= \big(G_{ij}\big)$ are given as a $4\times 1$ matrix of coupling strengths:
\renewcommand{\arraystretch}{1.25}\be\label{eq:couplings}
  {\cal G}  =  \bigg(
    -\frac{4 i e^2}{9 c_W^2},\quad
    -\frac{4 i e^2}{9 c_W^2},\quad
    -ig_s^2,\quad
    -ig_s^2 \bigg)^t\,.
\ee

The UFO format mimics this structure with the declaration of the model vertices as instances of the \verb+Vertex+ class in the file \verb+vertices.py+. Each vertex is implemented following its decomposition~\eqref{eq:decomp}, that is passed through five mandatory attributes (\verb+name+, \verb+particles+, \verb+color+, \verb+lorentz+ and \verb+couplings+). In the case of the four-squark vertex example considered, a possible instantiation is:
\begin{verbatim}
V_1 = Vertex(
  name      = 'V_1',
  particles = [ 
    P.suR,          P.suR,
    P.suR__tilde__, P.suR__tilde__
  ],
  color     = [
    'Identity(3,1)*Identity(4,2)',
    'Identity(4,1)*Identity(3,2)',
    'T(-1,1,3)*T(-1,2,4)',
    'T(-1,1,4)*T(-1,2,3)'
  ],
  lorentz   = [ L.SSSS1 ],
  couplings = {
     (0,0):C.GC_1,  (1,0):C.GC_1,
     (2,0):C.GC_2,  (3,0):C.GC_2
  }
)
\end{verbatim}
The first attribute {\tt name} defines the name given to the vertex (\verb+V_1+ in our example). The list of particles outgoing from the vertex is provided as an array of \verb+Particle+ objects through the \verb+particles+ attribute of the \verb+Vertex+ class. All employed particles must have been declared in the file \verb+particles.py+, and then imported in the file \verb+vertices.py+ prior to the declaration of any vertex as
\begin{verbatim}
import particles as P
\end{verbatim}
The four-squark example considered involves two incoming right-handed up squark (\verb+suR+) and two incoming right-handed up antisquarks (\verb+suR__tilde__+), the \verb+Particle+ objects \verb+suR+ and \verb+suR__tilde__+ being declared in {\tt particles.py} (as detailed in section~\ref{sec:particles}). The vertex decomposition~\eqref{eq:decomp} is finally provided through the \verb+color+, \verb+lorentz+ and \verb+couplings+ attributes of the \verb+Vertex+ class.

\renewcommand{\arraystretch}{1.4}%
\begin{table}
  \centering\setlength{\tabcolsep}{6pt}
  \begin{tabular}{rp{5.05cm}}
  UFO colour tensor & Description\\
  \hline
   \verb+1+                 & Trivial tensor (for non-coloured particles)\\
   \verb+Identity(2,1)+     & Kronecker delta $\delta^{\bar \imath_2}{}_{i_1}$, $\delta^{a_2a_1}$, or $\delta^{\bar \alpha_2}{}_{\alpha_1}$\\
   \verb+T(1,2,3)+          & Fundamental representation matrix
                                $(T^{a_1})^{\bar \imath_3}{}_{i_2}$\\
   \verb+f(1,2,3)+          & Antisymmetric structure constant $f^{a_1a_2a_3}$\\
   \verb+d(1,2,3)+          & Symmetric structure constant $d^{a_1a_2a_3}$\\
   \verb+Epsilon(1,2,3)+    & Fundamental Levi-Civita tensor
                                $\epsilon_{i_1i_2i_3}$\\
   \verb+EpsilonBar(1,2,3)+ & Antifundamental Levi-Civita tensor
                                $\epsilon^{{\bar \imath_1}{\bar \imath_2}{\bar \imath_3}}$\\
   \verb+T6(1,2,3)+         & Sextet representation matrix
                                $(T_6^{a_1})^{\bar \alpha_3}{}_{\alpha_2}$\\
   \verb+K6(1,2,3)+         & Sextet Clebsch-Gordan coefficient
                                $(K_6)^{{\bar \imath_2}{\bar \imath_3}}{}_{\alpha_1}$\\
   \verb+K6Bar(1,2,3)+      & Antisextet Clebsch-Gordan coefficient
                                $(\overline K_6)^{\bar \alpha_1}{}_{i_2i_3}$
  \end{tabular}
  \caption{Elementary colour tensors that can be used to construct the elements of the colour basis relevant for a given UFO vertex. Fundamental, sextet, antifundamental and antisextet colour indices are denoted as $i$, $\alpha$, $\bar \imath$ and $\bar\alpha$, whilst $a$ denotes an adjoint colour index.}
  \label{tab:color}
\end{table}
\renewcommand{\arraystretch}{1.}%

The \verb+color+ attribute refers to the array of elements $C_i^{a_1\ldots a_n}$ of the colour basis relevant to the vertex under consideration. Each entry in this array is a polynomial combination of the elementary colour tensors of table~\ref{tab:color}, and the arguments of each tensor are positive or negative integer numbers. Positive integers are used to associate a colour index with one of the particles incoming to the vertex, the exact value referring to the position of the particle in the list provided through the attribute \verb+particles+ of the \verb+Vertex+ class. Negative integers must appear exactly twice in a monomial, and they correspond to contracted (\ie\ summed over) indices. In the UFO conventions, the position of the first particle in the list \verb+particles+ corresponds to 1, in contrast to standard \python\ arrays. Moreover, it is up to users to verify the consistency between the colour structures appearing in a vertex definition and the representations of the particles entering this vertex, as programs processing UFO models may reject models in which the colour structures in a vertex do not match the colour representations of the particles.

Eq.~\eqref{eq:UFOex} shows that all the colour structures appearing in the four-squark vertex can be implemented by the sole use of Kronecker deltas (\verb+Identity+) and fundamental representation matrices of $SU(3)$ (\verb+T+). Consequently, the elements of the basis of \eqref{eq:colobasis} are implemented as 
\begin{eqnarray*}
 \delta^{\bar c_4}{}_{c_1}\delta^{\bar c_3}{}_{c_2} &\leadsto& \verb+'Identity(4,1)*Identity(3,2)'+\\
 \delta^{\bar c_3}{}_{c_1}\delta^{\bar c_4}{}_{c_2} &\leadsto& \verb+'Identity(3,1)*Identity(4,2)'+\\
 (T^a)^{\bar c_3}{}_{c_2} (T^a)^{\bar c_4}{}_{c_1}  &\leadsto& \verb+'T(-1,2,3)*T(-1,1,4)'+\\
 (T^a)^{\bar c_3}{}_{c_1} (T^a)^{\bar c_4}{}_{c_2} &\leadsto& \verb+'T(-1,1,3)*T(-1,2,4)'+
\end{eqnarray*}
as illustrated in the declaration of the vertex \verb+V_1+ above.

\renewcommand{\arraystretch}{1.4}%
\begin{table}
  \begin{tabular}{rp{5.32cm}}
  UFO spin tensor  & Description\\
  \hline
   \verb+Identity(1,2)+    & (Spinorial) Kronecker delta $\delta_{s_1s_2}$\\
   \verb+IdentityL(1,2)+   & (Lorentz) Kronecker delta $\delta^{\mu_1}_{\mu_2}$\\
   \verb+Gamma(1,2,3)+     & Dirac matrix $(\gamma^{\mu_1})_{s_2s_3}$\\
   \verb+Gamma5(1,2)+      & Fifth Dirac matrix $(\gamma^5)_{s_1s_2}$\\
   \verb+ProjM(1,2)+       & Left chirality projector $(\frac{1-\gamma_5}{2})_{s_1s_2}$\\
   \verb+ProjP(1,2)+       & Right chirality projector $(\frac{1+\gamma_5}{2})_{s_1s_2}$\\
   \verb+Sigma(1,2,3,4)+   & Sigma matrix $(\sigma^{\mu_1\mu_2})_{s_1s_2}$\\
   \verb+C(1,2)+           & Charge conjugation matrix $C_{s_1s_2}$\\
   \verb+Metric(1,2)+      & Minkowski metric $\eta^{\mu_1\mu_2}$\\
   \verb+P(1,i)+           & Incoming momentum of the $i^{\rm th}$ particle $p_i^{\mu_1}$\\
   \verb+Epsilon(1,2,3,4)+ & Levi-Civita tensor $\epsilon^{\mu_1\mu_2\mu_3\mu_4}$ (with $\epsilon_{0123}=-\epsilon^{0123}=1$)
  \end{tabular}
  \caption{Elementary spin tensors that can be used to construct the elements of the spin basis relevant to a given UFO vertex. Spin and Lorentz indices are respectively denoted as $s$ and $\mu$.
  } \label{tab:spins}
\end{table}
\renewcommand{\arraystretch}{1.}%

Similarly, all spin structures $L_j^{\ell_1\ldots\ell_n}(p_1,\ldots,p_n)$ relevant to a given vertex are collected into an array that is passed through the \verb+lorentz+ attribute of the \verb+Vertex+ class. The structures $L_j^{\ell_1\ldots\ell_n}$ are provided as \verb+Lorentz+ objects, instead of being directly implemented at the time of the vertex instantiation. These \verb+Lorentz+ objects are then defined in the file \verb+lorentz.py+, and they must therefore be imported prior to the declaration of any vertex in the file \verb+vertices.py+ file,
\begin{verbatim}
import lorentz as L
\end{verbatim}
A Lorentz object is instantiated (in the file \verb+lorentz.py+) as in the following two examples (the first one being the only one relevant for the considered four-squark interaction vertex),
\begin{verbatim}
SSSS1 = Lorentz(
  name      = 'SSSS1',
  spins     = [ 1, 1, 1, 1 ],
  structure = '1'
)

VVSS1 = Lorentz(
  name      = 'VVSS1',
  spins     = [ 3, 3, 1, 1 ],
  structure = 'Metric(1,2)'
)
\end{verbatim}
All three attributes of each \verb+Lorentz+ object are mandatory. The first of them (\verb+name+) indicates the name of the object, the second (\verb+spins+) the spins (in the $2s+1$ notation) of the particles entering the vertex and the last one (\verb+structure+) the structure itself, provided as a polynomial combination of the elementary tensors of table~\ref{tab:spins}. As in the colour case, the arguments of these tensors are positive and negative integers, the positive ones being associated with the particles incoming to the vertex (with the value referring to the position of the particle in the list \verb+spins+), and the negative ones appearing twice and corresponding to contracted indices (that are therefore summed over). In this context, squared momenta like $p_1^2$ can be written as \verb+P(-1,1)**2+. This is interpreted exactly as \verb+P(-1,1)*P(-1,1)+ and allows for concise expressions of  Lorentz structures in UFO.  Obviously, the \verb+-1+ index must appear only once in this case.

In the case of the four-scalar vertex~\eqref{eq:UFOex}, the only possible spin combination is the trivial one. This requires us to use the \verb+SSSS1+ object for the instantiation of the vertex \verb+V_1+. In contrast, the object \verb+VVSS1+ involves two vector bosons and two scalar particles (\cf\ the attribute \verb+spins+ of the object \verb+VVSS1+), and the structure of its interactions relates the two bosons through the Minkowski metric (\cf\ the attribute \verb+structure+ of the object \verb+VVSS1+). The corresponding UFO implementations for the structure of the \verb+Lorentz+ objects \verb+SSSS1+ and \verb+VVSS1+ are then
\begin{eqnarray*}
  1                 & \leadsto & \verb+'1'+\\
  \eta^{\mu_1\mu_2} & \leadsto & \verb+'Metric(1,2)'+
\end{eqnarray*}

The last attribute of the \verb+Vertex+ class is related to the coordinates $G_{ij}$ of the vertex in the colour $\otimes$ spin basis. They are provided in the form of a \python\ dictionary through the \verb+couplings+ attribute of the \verb+Vertex+ class. This dictionary relates the coordinate $(i,j)$, where $i$ and $j$ refer to a specific colour and spin basis element respectively, to the value of the corresponding coupling strength given as a \verb+Coupling+ object. The list of all \verb+Coupling+ objects necessary for the implementation of a given model is declared in the file \verb+couplings.py+, that must therefore be imported prior to the declaration of any vertex,
\begin{verbatim}
import couplings as C
\end{verbatim}
When a vertex is instantiated, only the non-vanishing coordinates have to be included. In the four-squark vertex considered, this therefore gives
\begin{eqnarray*}
  (0, 0) \leadsto \verb+C.GC_1+ &\qquad\qquad&
  (1, 0) \leadsto \verb+C.GC_1+\\%
  (2, 0) \leadsto \verb+C.GC_2+ &\qquad\qquad&
  (3, 0) \leadsto \verb+C.GC_2+
\end{eqnarray*}
where the integer counters follow this time a standard \python\ numbering for the elements of an array (the first element being thus associated with the index 0). The example above illustrates the fact that the four-squark vertex exemplified involves only two instances of the \verb+Coupling+ class (\verb+GC_1+ and \verb+GC_2+), as also depicted in~\eqref{eq:UFOex}.

The declaration of a \verb+Coupling+ object in the file \verb+couplings.py+ is very similar to that of an internal parameter declaration (see section~\ref{sec:parameters}). For the two couplings \verb+GC_1+ and \verb+GC_2+ necessary for the four-squark vertex considered, this gives
\begin{verbatim}
GC_1 = Coupling(
 name  = 'GC_1',
 value = '-(4*ee**2*complex(0,1))/(9.*cw**2)',
 order = {'QED':2}
)

GC_2 = Coupling(
 name  = 'GC_2',
 value = '-(complex(0,1)*G**2)',
 order = {'QCD':2}
)
\end{verbatim}
An instance of a \verb+Coupling+ object is declared with three mandatory arguments, namely its name (\verb+name+), the algebraic coupling definition that could depend on any of the model parameters (\verb+value+), and a so-called coupling order provided in the form of a \python\ dictionary (\verb+order+). In the case of the four-squark example considered, the coupling strengths appearing in \eqref{eq:couplings} are directly provided as valid \python\ algebraic expressions.  The last attribute of a \verb+Coupling+ object, \verb+order+, is a \python\ dictionary that allows users to tag certain couplings of the model with one or more strings (\ie\ tags) to which a positive integer number is associated. In the examples above, the tags \verb+QED+ and \verb+QCD+ are associated with two physical quantities, the typical strength of the electroweak and strong interactions. The couplings \verb+GC_1+ and \verb+GC_2+ are hence flagged as couplings with strengths proportional to two powers of the electromagnetic and strong coupling, respectively, as the integer 2 is attached with each of the two tags involved. It is not mandatory to use tags that actually refer to a physical interaction. For instance, in the vector-like quark UFO implementation of~\cite{Fuks:2016ftf}, a \verb+VLQ+ coupling order is introduced in order to identify all vertices suppressed by the mixing of a vector-like and a SM quark (which is achieved by setting \verb+order = {'VLQ:1'}+ in the relevant coupling declarations).

This coupling-order feature allows users to filter not only vertices, but also the resulting Feynman diagrams. This is generally achieved in practice through the introduction of criteria depending on the type of interactions involved in a vertex or a Feynman diagram. For instance, the \verb+QED+ and \verb+QCD+ tags introduced in the definition of the \verb+GC_1+ and \verb+GC_2+ couplings could allow users to neglect (subdominant) electroweak diagrams relative to (dominant) QCD diagrams (as numerically $\alpha_s^2 \sim \alpha$) when deriving the list of diagrams relevant to a specific hadron-collider process.%\footnote{Such a QCD/electroweak diagram filtering is generally not relevant for lepton colliders.} 
Moreover, in the vector-like quark example briefly mentioned above, users could enforce the list of relevant diagrams to include at most one mixing suppression.

The tags that can be used for the instantiation of the different \verb+Coupling+ objects have to be declared in the file \verb+coupling_orders.py+, each tag (or coupling order) being implemented as an instance of the class \verb+CouplingOrder+.  In our supersymmetric example (taken from~\cite{Frixione:2019fxg}), the model contains two independent classes of interactions that are named \verb+QED+ (for interactions proportional to the electromagnetic coupling $e$, and therefore  the weak coupling $g=e/s_W$ with $s_W$ being the sine of the electroweak mixing angle, or any of the model’s Yukawa or supersymmetry-breaking multiscalar interactions) and \verb+QCD+ (for QCD interactions). These tags are declared as
\begin{verbatim}
QCD = CouplingOrder(
  name                   = 'QCD',
  expansion_order        = 99,
  hierarchy              = 1,
  perturbative_expansion = 1
)

QED = CouplingOrder(
  name            = 'QED',
  expansion_order = 99,
  hierarchy       = 2
)
\end{verbatim}
Whereas the examples above refer to coupling orders associated with physical interactions, they can easily be generalised to any other class of tags.

In the above example, the two \verb+CouplingOrder+ objects are instantiated by fixing three mandatory attributes (\verb+name+, \verb+expansion_order+ and \verb+hierarchy+), together with one optional attribute for the \verb+QCD+ coupling order (\verb+perturbative_expansion+). The first of the mandatory arguments, \verb+name+, contains the name of the coupling order, that also consists of the tag that can be used for instantiation of \verb+Coupling+ objects. The second attribute, \verb+expansion_order+, refers to the maximum power of the interaction that could appear in a single amplitude. This is particularly relevant for effective field theories in which amplitudes must be truncated to some power of the high effective scale. In the above examples, this attribute is fixed twice to \verb+99+, which effectively indicates that there is no limit. The last mandatory attribute, \verb+hierarchy+, allows users to order the couplings according to their relative magnitude. In the above example, we enforce such a hierarchy, and we impose $\alpha_s^2 \sim \alpha$. This is achieved by assigning to the coupling orders \verb+QCD+ and \verb+QED+ the hierarchies 1 and 2 respectively. Such a piece of information is relevant for the implementation of diagram filters for a given process,\footnote{In most public UFO models, the tag {\tt QED} is traditionally associated with both the usual QED interactions and all electroweak interactions. Consequently, in the case of the top-quark Yukawa coupling (that is thus a coupling of {\tt QED} order), the {\tt QCD}/{\tt QED} hierarchy mentioned in the text is not valid. The {\tt QED} tag is however anyway very useful for diagram filtering purpose.} allowing users to select given contributions to an amplitude according to the type of contributing interactions. Setting the attribute \verb+hierarchy+ to~0 indicates that the corresponding coupling order plays only a role of a tag, and that there is no connection to the relative magnitude of the associated coupling orders. In the vector-like quark model introduced above, the attribute \verb+hierarchy+ of the coupling order \verb+VLQ+ is set to 0. This coupling order can hence be used to enforce the maximum number of suppression factors due to VLQ-SM mixing that can appear in a diagram, regardless of the nature of the fundamental (\verb+QED+ or \verb+QCD+) interactions involved.

Finally, the \verb+perturbative_expansion+ attribute of the \verb+QCD+ coupling order is set to \verb+1+ in the above example. This implies that the UFO model contains all ingredients necessary for NLO calculations in QCD (see section~\ref{sec:ufo_nlo}). If the attribute is unspecified, a default value of \verb+perturbative_expansion = 0+ is assumed, which implies that only LO calculations in this coupling are supported.

\subsection{The function library}\label{sec:fct}
The last mandatory component of a UFO model is the file \verb+function_library.py+. It includes user-defined functions declared as instances of the \verb+Function+ class. The UFO format supports functions that can be defined within a single line in \python, the so-called \python\ lambda functions. \python\ lambda functions offer the advantage of being easily translatable into other programming languages, and they are consequently the only functions that can be declared within the library defined in the file \verb+function_library.py+.

A \verb+Function+ object is defined through three mandatory attributes and two optional attributes. The mandatory attributes consist of the name of the function (\verb+name+), its arguments specified as a tuple of strings (\verb+arguments+), and the expression of the function itself given in terms of its arguments. The latter must be provided as a string that represents a valid \python\ expression, and it is given as the value of the attribute \verb+expression+. By default, all the arguments of the functions are considered to be complex numbers. This behaviour can however be superseded by providing a tuple of strings through the optional attribute \verb+argstype+ of the \verb+Function+ class, which allows users to specify the type of the different arguments of the function. The supported types are real numbers (\verb+real+), complex numbers (\verb+complex+), and arrays of real or complex numbers (\verb+real[n]+ or \verb+complex[n]+ for an \verb+n+-dimensional array, respectively, where \verb+n+ is an integer). The two tuples provided through the attributes \verb+arguments+ and \verb+argstype+ must have the same size. Similarly, the type of the result of the function is a complex number by default, but this behaviour can be modified by specifying the \verb+type+ attribute of the instantiated function, that can take the self-explanatory values \verb+real+, \verb+complex+, \verb+real[n]+ and \verb+complex[n]+ (with \verb+n+ being an integer).

Several functions are shipped by default with any UFO model. This includes in particular a series of mathematical functions for which the \python\ module \verb+cmath+ is insufficient. First, the \verb+function_library.py+ file contains a set of tools that facilitate the treatment of complex quantities (the real part of a complex number \verb+re+, its imaginary part \verb+im+, and the complex conjugation operation \verb+complexconjugate+). Second, several trigonometric and cyclometric functions are implemented (the cotangent \verb+cot+, the secant \verb+sec+ and the cosecant \verb+csc+ functions, together with their arcsecant \verb+asec+ and arccosecant \verb+acsc+ counterparts). 

As an illustration, we provide below a function returning the real part of a complex number, as well as a function associated with the secant of a complex number. These could implemented as
\begin{verbatim}
re = Function(
  name       = 're',
  arguments  = ('z'),
  expression = 'z.real'
)

sec = Function(
  name       = 'sec',
  arguments  = ('z'),
  expression = '1./cmath.cos(z.real)'
)
\end{verbatim}

Furthermore, the file \verb+function_library.py+ of a UFO model includes a generalised version of the Heaviside step function \verb+theta_function+. It allows users to make use of a variety of piecewise functions involving a single condition, and it is implemented as a \verb+Function+ object relying on the one-line \python\ if/else statement,
\begin{verbatim}
theta_function = Function(
  name       = 'theta_function',
  arguments  = ('x','y','z'),
  expression = 'y if x else z'
)
\end{verbatim}
With this generalised function, the familiar one-para\-me\-ter Heaviside function centred on $x_0=23$,
\be
  \Theta(x-23) = \left\{\begin{array}{l}
    1 \qquad \text{if~}x\geq 23.\\
    0 \qquad \text{otherwise}\end{array}\right.\,,
\ee
can be used through
\begin{verbatim}
  theta_function(x>=23., 1., 0.)
\end{verbatim}

The definition of \verb+Function+ objects also supports the use of the model parameters instantiated in the file \verb+parameters.py+, as well as that of other functions. This therefore allows for more complex expressions to be defined in steps. As a complete example, we consider an elastic atomic form factor $G_{\rm el}(t)$~\cite{Bjorken:2009mm},
\renewcommand{\arraystretch}{1.85}\be
  G_{\rm el}(t) =  \left( \frac{Z_{\rm nuc.}}{1+ t/d_{\rm nuc.}} \frac{a^2_{\rm nuc.} t}{1 + a_{\rm nuc.}^2 t} \right)^2\,,
\ee
which depends on the three nuclear physics parameters $Z_{\rm nuc.}$, $a_{\rm nuc.}$ and $d_{\rm nuc.}$.  Its implementation as a \verb+Function+ object \verb+Gel+ reads
\begin{verbatim}
Gel = Function(
  name       = 'Gel',
  type       = 'real',
  arguments  = ('t'),
  argstype   = ('real'),
  expression = '(Z_nuc/(1+t/d_nuc)*
      a_nuc**2*t/(1+a_nuc**2*t))**2'
) 
\end{verbatim}
where \verb+Z_nuc+, \verb+a_nuc+ and \verb+d_nuc+ are parameters of the model defined in the file \verb+parameters.py+. In this function, we imposed the arguments to be real quantities (\verb+real+), instead of complex ones (\verb+complex+, that is also the default) or real and complex arrays (for example \verb+real[4]+ or \verb+complex[4]+ for the four-dimensional case). Moreover, the output of the function is defined to be a real number, instead of a complex one, which is the default. We note that this output is not allowed to be a list. 

The complete form factor in our example should also include an inelastic part $G_{\rm in}(t)$~\cite{Bjorken:2009mm, Jodlowski:2019ycu}. The latter can be defined similarly as its elastic counterpart, namely as another \verb+Function+ object \verb+Gin+.  The full implementation of both contributions as a single \verb+Function+ object \verb+FF+ is thus
\begin{verbatim}
FF = Function(
  name       = 'FF',
  arguments  = ('t'),
  argstype   = ('real'),
  expression = 'math.sqrt(Gel(t) + Gin(t))'
)
\end{verbatim}
in which we illustrate how a given \verb+Function+ object could call other \verb+Function+ objects and standard \python\ methods.

\section{Optional components}
\label{sec:ufo_optional}
In this section, we describe optional files that can be included in a UFO model. These files allow users to provide additional information about a model, and/or equip a UFO with non-standard practical methods and functions. One of these files (\verb+write_param_card.py+) defines a writer of the external model parameters in an SLHA-like text file. This file was a mandatory component in the first version of the UFO format~\cite{Degrande:2011ua} for the sole reason that all files defining a model were mandatory. As it is not strictly necessary from the point of view of the information defining a model, we benefit from the possibility of having optional files in version 2 of the UFO to update its nature. The other optional files are new and were introduced after the original release of the UFO format. They are related to the addition of custom propagators for specific particles of the model as introduced in~\cite{Christensen:2013aua} (in the file \verb+propagators.py+), detail how to provide information about particle decay widths following~\cite{Alwall:2014bza} (in the file \verb+decays.py+) and about the renormalisation group running of the model's parameters as defined in~\cite{Aoude:2022aro} (in the file \verb+running.py+), and enable the usage of form factors in a UFO model, which we document below for the first time.  Moreover, it is now also possible to directly add custom {\sc Fortran} and {\sc C++} functions in a UFO model. These functions are defined in folders \verb+Fortran+ and \verb+Cpp+ respectively, and they can be called in any algebraic expression introduced in the other files of the model. This possibility is briefly discussed in sections~\ref{sec:formfactors} and \ref{subsec:complex_mass_scheme}.

\subsection{Outputting the values of the model parameters}
The file \verb+write_param_card.py+ includes routines that write all external model parameters, together with their numerical value, into a text file following an SLHA-like format~\cite{Skands:2003cj,Allanach:2008qq}. In the output file, the parameters and their values are organised in Les Houches blocks and counters, as specified by the user with the parameter declarations implemented in the file \verb+parameters.py+ (\cf\ the \verb+lhablock+ and \verb+lhacode+ attributes of the different parameters; see section~\ref{sec:parameters}). 

The output file, named \verb+param_card.dat+, is generated by issuing in a shell the command
\begin{verbatim}python write_param_card.py\end{verbatim}
In addition to the model external parameters and their value, the output file includes \verb+QNUMBERS+ blocks~\cite{Alwall:2007mw} with information on the quantum numbers of all the particles of the model, as well as all particle masses and decay widths regardless of their external/internal nature. Examples of such a \verb+write_param_card.py+ file can be obtained from the model database of \feynrules.\footnote{See the webpage \url{https://feynrules.irmp.ucl.ac.be/wiki/ModelDatabaseMainPage}.}

\subsection{Form factors}\label{sec:formfactors}

The standard UFO decomposition of an interaction vertex in a colour $\otimes$ spin space of \eqref{eq:decomp} does not always suffice to properly describe an interaction. In some models, it is indeed convenient to have couplings that depend on phase space (therefore including so-called form factors), as for instance in effective theories or empirical descriptions of interactions (\eg\ as for atomic form factors at low energy or for neutrino-nucleus interactions). The extension of the UFO format described in this section adopts the decomposition~\eqref{eq:decomp} by adding extra scalar functions $\begin{cal}F\end{cal}_{j}\,(p_1,\ldots,p_n)$ that depend on the four-momenta of the particles incoming to the vertex,
\be\bsp
  &\quad {\cal V}^{a_1\ldots a_n, \ell_1\ldots\ell_n} =
    \sum_{i,j}C_i^{a_1\ldots a_n} \ G_{ij}\\
    &\qquad\quad\times  \begin{cal}F\end{cal}_{j}\,(p_1,\ldots,p_n)\ L_j^{\ell_1\ldots\ell_n}(p_1,\ldots,p_n) \ .
\esp\label{eq:generic_vertex2}\ee
This expression shows that in the UFO conventions, the form factors $\begin{cal}F\end{cal}_{j}$ impact the spin dependence of the interaction vertices, while they leave the colour structure unaffected. 

In practice, they are implemented as a modification of the relevant spin structure of the vertices declared in the file \verb+lorentz.py+, following the replacement
\be\bsp
  & L_j^{\ell_1\ldots\ell_n}(p_1,\ldots,p_n) \to \\ & \qquad\qquad \mathcal{F}_{j}\,(p_1,\ldots,p_n)\ L_j^{\ell_1\ldots\ell_n}(p_1,\ldots,p_n)\,.
\esp\ee
This amounts to allow the value of the \verb+structure+ attribute of a \verb+Lorentz+ object to make use of functions defined in the file \verb+function_library.py+ (see section~\ref{sec:fct}), and of parameters defined in the file \verb+parameters.py+ (see section~\ref{sec:parameters}). This obviously requires to import the list of parameters and the set of relevant functions in the preamble of the file {\tt lorentz.py}. 

As an example, we consider the case of a form factor given by $m_W/E$, where $m_W$ stands for the mass of the $W$ boson (represented below by the \verb+Parameter+ object \verb+MW+) and $E$ is the energy scale relevant for the associated process. Such a form factor could be defined in the file {\tt function\_library.py} through a \verb+Function+ object \verb+AAA+,
\begin{verbatim}
AAA = Function(
  name       = 'AAA',
  type       = 'float',
  arguments  = ('E2'),
  argstype   = ('float'),
  expression = 'MW/cmath.sqrt(E2)'
)
\end{verbatim}
This form factor can then be used in the declaration of a spin structure relevant, for instance, for a vertex involving two vector bosons (of momenta $p_1$ and $p_2$ and associated Lorentz indices $\mu_1$ and $\mu_2$), and one scalar state (of momentum $p_3$), 
\be
  \mathcal{F}\,(p_1,p_2,p_3)\ L^{\mu_1\mu_2}(p_1,p_2,p_3) = \frac{m_W\ \  \eta^{\mu_1\mu_2} }{\sqrt{(p_1+p_2)^2}} \,.
\ee
In this expression, the energy scale $E$ appearing in the form factor is identified by $E^2 \equiv (p_1+p_2)^2$. This could be implemented as a \verb+Lorentz+ object \verb+VVS1+ as 
\begin{verbatim}
VVS1 = Lorentz(
  name      = 'VVS1',
  spins     = [ 3, 3, 1 ],
  structure = 'AAA((P(-1,1)+P(-1,2))**2)
     * Metric(1,2)' 
)
\end{verbatim}
following the notation introduced in table~\ref{tab:spins}. In particular, we recall that negative indices are summed over, and that squares of four-vectors are allowed.

For more complicated form factor expressions, users have the possibility to nest the definition of several functions, as shown in the example given in section~\ref{sec:fct} or in \cite{Bonciani:2022jmb, Becchetti:2020wof}. For extreme cases, form factors can be provided externally, through {\sc Fortran} or  {\sc C++} functions as in the tau-lepton decay module of \cite{Hagiwara:2012vz}. Such a construction should however be avoided as much as possible as it breaks the spirit of portability of UFO models. It may however be sometimes the only choice. In this case, we encourage authors to provide both {\sc Fortran} and {\sc C++} routines for their form factors, and implement them in the {\sc Fortran} file \verb+Fortran/functions.f+ (or \verb+Fortran/functions.f90+) and  {\sc C++} header and source files \verb+Cpp/functions.h+ and \verb+Cpp/functions.cpp+ respectively.

\subsection{Particle propagators}\label{sec:propagators}

In general, the propagator of a particle can be inferred from its spin, and so it is usually redundant to define propagators explicitly for each particle. There may be cases, however, where it is useful to have the possibility to redefine the propagator of a certain class of particles. This includes, for example, theories with non-standard kinetic terms, implementations featuring non-propagating auxiliary particles (in which case the propagator is simply a product of Kronecker delta functions without any momentum dependence), and models relevant for particles with a spin value $s\geq 3/2$ for which the conventions are not unique.

For a few years already, the UFO format has allowed the user to define new propagators as instances of the class {\tt Propagator}, and in a given model implementation all these custom propagator definitions must be collected in the optional file {\tt propagators.py}~\cite{Christensen:2013aua}. The instantiation of a \verb+Propagator+ object follows similar conventions as for any other UFO object, as exemplified below with the case of a massless gauge boson propagator in the Feynman gauge (instantiated as \verb+V0+),
\begin{verbatim}
V0 = Propagator(
  name        = "V0",
  numerator   = "-1 * Metric(1, 2)",
  denominator = "P(-1, id)**2"
)
\end{verbatim}
This declaration includes the two mandatory attributes of the \verb+Propagator+ class (\verb+name+ and \verb+numerator+), as well as the only possible optional attribute (\verb+denominator+). The attribute \verb+name+ provides a way to identify a given propagator object, whereas the attribute \verb+numerator+ includes an analytical expression for the numerator of the propagator, a global factor $i$ excluded. The optional attribute \verb+denominator+ then allows users to provide an analytical expression of the denominator. If unspecified, the Feynman propagator denominator $(p^2-m^2+im\Gamma)$ is assumed for a particle of mass $m$, width $\Gamma$ and four-momentum $p_\mu$.

\begin{table}
\renewcommand{\arraystretch}{1.35}
\setlength{\tabcolsep}{6pt}
\begin{tabular}{r p{5.5cm}}
  UFO expression & Description \\ \hline
  \verb+P(1,id)+ & Momentum of the propagating particle in a direction aligned with the incoming momentum flow\\
  \verb+P(2,id)+ & Momentum of the propagating particle in a direction aligned with the outgoing momentum flow \\
  \verb+Mass(id)+ & Mass of the propagating particle\\
  \verb+Width(id)+ &  Width of the propagating particle \\
  \verb+OverMass2(id)+ & $1/M^2$ for massive particle, and 0 otherwise\\
  \verb+PSlash(1,2,id)+ & $\slashed{p}_{s_1s_2}$ where $p$ is the momentum of the propagating particle\\&\\
\end{tabular}
  \caption{Lorentz objects that can be used for the definition of the numerator and denominator of custom particle propagators. \label{tab:lorentz}}
\end{table}

The analytical expressions to be provided for the propagator numerators and denominators rely on the UFO conventions detailed in section~\ref{sec:interactions} (and in table~\ref{tab:spins} in particular), as well as on several additional quantities that are introduced in table~\ref{tab:lorentz}. For non-scalar propagators, the numerator involves non-contracted (spin and/or Lorentz) indices that are referred to as `1' and `2' in the implementation. These respectively correspond to the incoming and outgoing directions. For non-fer\-mi\-o\-nic propagators, these directions are arbitrary, whereas for fermionic propagators they are crucial and must be defined from the `fermion flow' associated with the corresponding diagrams~\cite{Denner:1992me}. In the case of a spin-2 particle, the `51' and `52' indices are additionally introduced for the second pair of Lorentz indices attached to the propagating state. 

As shown in the example above and in table~\ref{tab:lorentz}, the flag \verb+id+ is used as the unique identifier for the propagating particle. For instance, the momentum of the propagating particle can be represented by \verb+P(1,id)+ and \verb+P(2,id)+ in the incoming and outgoing cases respectively, whereas for a fermionic propagator (of momentum $p$), \verb+PSlash(1,2,id)+ would refer to the quantity $(\slashed{p})_{s_1s_2}$ in spin space. Moreover, the mass and width of the propagating particle are identified as \verb+Mass(id)+ and \verb+Width(id)+ respectively, and the additional quantity \verb+OverMass2(id)+ corresponds to $1/M^2$ for a massive particle and 0 otherwise. Finally, we emphasise that as for Lorentz structure definitions, repeated negative indices are summed over.

In order to link custom propagators to particles, the {\tt Particle} class is equipped with an optional attribute {\tt propagator}. It allows users to refer to the specific propagator to employ through its name as defined in the file {\tt propagators.py}. For example, a massless spin-$1$ particle with a custom propagator as given in the above example could be defined by
\begin{verbatim}
photon = Particle(
  pdg_code    = 22,
  name        = 'photon',
  antiname    = 'photon',
  spin        = 3,
  color       = 0,
  mass        = Param.ZERO,
  width       = Param.ZERO,
  propagator  = propagators.V0
  texname     = '\gamma',
  antitexname = '\gamma',
  charge      = 0
)
\end{verbatim}
where the preamble of the file \verb+particles.py+ is assumed to include the instruction
\begin{verbatim}
import propagators
\end{verbatim}
In the case a \verb+Particle+ object is instantiated without any value for the attribute {\tt propagator} (as in most existing UFO models), default propagators are assumed. As non trivial and existing examples, we refer to \cite{Christensen:2013aua} and \cite{Deandrea:2021vje}. They respectively address models featuring particles with spin $s\geq 3/2$, and models in which running width effects are incorporated in the particle propagators.

Finally, some models are such that it is impossible to simultaneously diagonalise both the mass and width matrices. In this case, `matrix' propagators are in order~\cite{Cacciapaglia:2009ic}. We mention as a side note that the UFO format is compliant with such a structure through the implementation of a set of two-point vertices that emulate each off-diagonal entry of the propagator matrix.

\subsection{Particle decays}

Many applications of calculations involving massive unstable particles require the evaluation of the total and partial decay widths of all particles of the model, together with the estimation of the decay channels that are kinematically allowed. This task is highly dependent on the mass spectrum of the model, and it requires a re-evaluation of the widths for each choice of external parameters. In order to provide a simple solution to this problem, the UFO format allows users to input analytical formulas for LO two-body decay rates associated with the particles of the model. These are all collected inside the file {\tt decays.py}~\cite{Alwall:2014bza}. As two-body decays might sometimes be insufficient (when for instance higher-multiplicity decays are the dominant decay modes or when higher-order corrections are important), it is up to the code using the UFO model to decide how (and if) they should include such extra contributions in their computations.

In the special case of a two-body decay of a particle of mass $M$ to two particles of masses $m_1$ and $m_2$, Lorentz invariance implies that the matrix element relevant for the calculation of a partial width $\Gamma$ can only depend on the masses of the external particles, and we can write
\begin{equation}\label{eq:gamma}
  \Gamma = \frac{\sqrt{\lambda(M^2,m_1^2,m_2^2)}\,}{16\,\pi\,S\,|M|^3}\ |\mathcal{M}|^2 \,,
\end{equation}
where $S$ denotes the phase-space symmetry factor, the function $\lambda(M^2,m_1^2,m_2^2)= (M^2-m_1^2-m_2^2)^2-4m_1^2m_2^2$ is the usual K\"all\'en function, and $|\mathcal{M}|^2$ stands for the average squared matrix element associated with the decay mode considered. The matrix element of this two-body decay only receives contributions from a single three-point vertex $\mathcal{V}$, so that it can be written as
\begin{equation}\label{eq:master}
|\mathcal{M}|^2 = \mathcal{V}_{\mu_1\mu_2\mu_3}^{a_1a_2a_3}\,\mathcal{P}^{\mu_1\mu'_1}_1  \mathcal{P}^{\mu_2\mu'_2}_2\,\mathcal{P}^{\mu_3\mu'_3}_3\,(\mathcal{V}^*)_{\mu'_1\mu'_2\mu'_3}^{a_1a_2a_3}\,,
\end{equation}
where the colour and spin indices of the particle $i$ are generically denoted by $a_i$ and $\mu_i^{(')}$ respectively. In addition, we have introduced the polarisation tensor of the particle $i$, $\mathcal{P}_i$, that depends on its spin and its mass.

The content of the file \verb+decays.py+ contains declarations of instances of the class {\tt Decay}. Each instance of this class can be thought of as a collection of LO analytic formulas of two-body partial widths of a given state (obtained from \eqref{eq:gamma} and \eqref{eq:master}). For example, the two-body partial widths of the Higgs boson in the Standard Model could be represented as
\begin{lstlisting}
Decay_H = Decay(
  name           = 'Decay_H',
  particle       = P.H,
  partial_widths = {
   (P.W__minus__, P.W__plus__): '$\Gamma_{WW}$',
   (P.Z, P.Z): '$\Gamma_{ZZ}$',
   (P.b, P.b__tilde__): '$\Gamma_{b\bar b}$',
   (P.ta__minus__, P.ta__plus__):'$\Gamma_{\tau\tau}$',
   (P.t,P.t__tilde__): '$\Gamma_{t\bar t}$'
  }
)
\end{lstlisting}
where $\Gamma_{XY}$ schematically represent the analytic formula of the partial width of the Higgs boson associated with the decay mode $H\to XY$. The syntax to be used to write these analytic formulas is identical to that introduced in the previous sections. In the above example we assume that the first two generations of fermions are massless. All possible LO two-body decays have been included, even if some of them are kinematically forbidden. The analytic formula for the two-body decays of a Higgs boson into a pair of top quarks or weak bosons are hence also present (even if not kinematically allowed for a light Higgs boson). It is then up to the high-energy physics tool to filter out at run time the kinematically allowed channels (that depend on the chosen set of external parameters), and to combine them consistently into the total width and branching ratios for a given particle. When implementing the file \verb+decays.py+, it is strongly recommended to include {\it all} two-body decay channels for {\it all} the particles of the model, kinematically allowed or not, in order to prevent the code that relies on this option of the UFO format from producing incorrect results for some benchmark scenarios and correct ones for others.

The example of the Higgs boson is also a case where tree-level two-body decays are not sufficient for an accurate calculation of the total width of the particle. We should indeed include important contributions arising both from loop-induced Feynman diagrams and from three-body decays with off-shell effects, and the explicit choice of the renormalisation scale is known to impact the results strongly. We emphasise that it is not the role of the UFO model to check whether the provided formulas are enough to compute reliably the total width of a particle. The inclusion of the file \verb+decays.py+ in a UFO model instead only provides some analytical formula to facilitate the approximate evaluation of the particle widths.

\subsection{Renormalisation group running effects}

In many practical applications in particle physics, the free parameters of the Lagrangian are provided at a given input scale that could be quite different from the natural scales relevant to the physics process considered. One possibility to increase the precision of the predictions can therefore be to include renormalisation group (RG) running effects, which amounts to re-evalu\-a\-ting couplings and/or masses of the model at a specific scale. The UFO format has been extended~\cite{Aoude:2022aro} so that information on RG running could be provided within the optional file \verb+running.py+. Following the general UFO philosophy, the UFO format only contains information on the running of the model's parameters, and it does not provide any method allowing to handle it numerically. It is hence up to the high-energy physics software employed to handle this, and/or to rely on any existing external tools like those introduced in \cite{Staub:2013tta, Goodsell:2017pdq, Sartore:2020gou, DiNoi:2022ejg}.

In full generality, the RG equations associated with the model parameters $\{C\} = \{c_1, c_2, \ldots\}$ can be written as
\begin{equation}
  \frac{{\rm d} c_i(\mu)}{{\rm d}\log\mu} = \gamma^{(1)}_{ij}\ c_j(\mu) + \gamma^{(2)}_{ijk}\ \,c_j(\mu)c_k(\mu)+ \dots \,, \label{eq:rge}
\end{equation} 
where the anomalous dimension matrix $\gamma$ has been decomposed into a part involving a single other parameter ($\gamma^{(1)}_{ij}$), a part involving two other parameters ($\gamma^{(2)}_{ijk}$), and so on. The file \texttt{running.py} contains the values of the (non-zero) elements of the various $\gamma$ tensors appearing in the right-hand side of \eqref{eq:rge}. Several of these elements can be defined simultaneously, provided that they correspond to the same analytical expression, which allows for an economical implementation. Consequently, users have the possibility to declare one \verb+Running+ object for each unique value of the elements of the various $\gamma$ tensors appearing in an RG equation given by \eqref{eq:rge}, instead of one \verb+Running+ object per summand appearing in its right-hand side.

In practice, a \verb+Running+ object is defined as:
\begin{verbatim}
RGE_1 = Running(
  name        = 'RGE_1',
  value       = '2./(3.*cmath.pi)',
  run_objects = [
    [P.c1, P.c2, P.gs], 
    [P.c3, P.gs]
  ]
)
\end{verbatim}
This declaration relies on three mandatory attributes. The first of them is the name of the object that is provided as a string ({\tt name}), while the second attribute (\verb+value+) refers to the analytical formula associated with the elements of the $\gamma$ tensors considered. This formula has to be provided as a valid \python\ expression that follows the same technical limitations as those inherent to the \verb+value+ attribute of the classes \verb+Parameter+ and \verb+Coupling+ (see sections~\ref{sec:parameters} and \ref{sec:interactions}). Moreover, this expression should not depend on any running parameter. 

The value of the {\tt run\_objects} attribute contains a list in which each element is a list of external parameters (that must therefore be declared as instances of the \verb+Parameter+ class), with the exceptions of the standard QCD and QED couplings $\alpha_s$, $\alpha$, $g_s$ and $e$ that could be used despite their external/internal nature. For each entry in the primary list, the first parameter corresponds to the parameter appearing on the left-hand side of \eqref{eq:rge}, while all the other entries correspond to the parameters appearing on the right-hand side of that equation. In addition, a given parameter can be repeated as many times as needed to obtain a dependency on a specific power of it. The above example would correspond to
\begin{equation}\begin{split}
  \frac{{\rm d} c_1(\mu)}{{\rm d}\log\mu} =& \frac{2}{3\pi}\ g_s c_2 + \dots\,, \\
  \frac{{\rm d} c_3(\mu)}{{\rm d}\log\mu} =& \frac{2}{3\pi}\ g_s + \dots\,,
\end{split} \end{equation}
where the dots refer to terms not captured by the declaration of the \verb+RGE_1+ object above.

\section{Features pertaining to NLO}
\label{sec:ufo_nlo}
% !TEX root = UFO_Paper.tex

Up to this point, we have presented many aspects of the UFO format that provide the necessary information to generate tree-level matrix elements for arbitrary processes within a model. In principle, this information is also sufficient to produce matrix elements that include corrections associated with an arbitrary number of loops. In practice, however, a number of additional ingredients are necessary, mainly in the form of \emph{process-independent} counterterms whose derivation is often quite involved. They should therefore ideally be supplied along with the rest of the tree-level information included in a UFO model. In this section, we describe the standard according to which this additional information is provided. We stress that this extended UFO format is not bound to one-loop corrections. However current applications only involve one-loop automatic matrix element generation, and it is therefore the case that drove our choice of syntax.

The counterterms provided in a UFO model containing the necessary information for performing computations at NLO accuracy (afterwards referred to as an NLO UFO) come in two distinct categories called $R_2$ and UV. The $R_2$ category contains \emph{rational terms of the second kind}. They originate from the need of recovering contributions from the $d$-dimensional part of one-loop numerators that are typically computed in four dimensions by most numerical approaches~\cite{Draggiotis:2009yb, Garzelli:2009is,Garzelli:2010qm,Shao:2011tg,Pittau:2011qp,Shao:2012ja,Page:2013xla,Chen:2019fhs}. The UV category implements the ultraviolet renormalisation of the model. This requires an analysis of the loop corrections to the vertices and two-point functions of the model, together with physically motivated choices made by the model builder (\eg\ renormalisation conditions). For this reason, such counterterms are also best computed once and for all and specified in the UFO model. As it is further discussed below, an NLO UFO model is therefore suitable for one (or at most a few) particular renormalisation scheme(s).

We now present the standardised format in which these {$R_2$} and {\sc UV} counterterms are provided in an NLO UFO model, and that is already used by many one-loop providers (OLPs). This is however only achieved after first briefly introducing information relevant for one-loop matrix element computations in sections~\ref{subsec:counterterms} (UV and $R_2$ counterterms) and \ref{subsec:complex_mass_scheme} (the complex mass scheme). In section~\ref{subsec:ew}, we focus on electroweak corrections in the Sudakov approximation and detail why additional pieces of information must be provided. As mentioned in section~\ref{sec:generalstructure}, several additional files must be supplied in an NLO UFO model. We provide details about those files in section~\ref{subsec:ctufo}. They consist of the file \verb+CT_vertices.py+\footnote{The letters `{\tt CT}' appearing at the beginning of the filename refer to the word `counterterm', although the information included in the UFO files described in the present section does not only concern counterterms {\it stricto sensu}.} that allows for the instantiation of all \verb+CTVertex+ objects included in the model, the file \verb+CT_couplings.py+ that allows for the declaration of \verb+Coupling+ objects used in the counterterms, and the file \verb+CT_parameters.py+ that includes instantiation of all \verb+CTParameter+ objects needed in the counterterm couplings. 

Additional information not available in any other UFO file is needed for making it possible to automatically calculate electroweak Sudakov corrections. This consists of the eigenvalues of various electroweak operators, and on how the components of the physical fields in a theory are gathered into $SU(2)_L$ multiplets. This is provided in the file \verb+CT_ewcasimirs.py+ through the declaration of \verb+EWOperator+ objects. 

The objects specific to UFO NLO models can easily be accessed through generic lists included in the file \verb+object_library.py+. A first list \verb+all_CTvertices+ collects all declared \verb+CTVertex+ objects, while a second list \verb+all_CTparameters+ is dedicated to the \verb+CTParameter+ objects declared by the user. On the other hand, all counterterm couplings (declared as standard \verb+Coupling+ objects) are available together with the other couplings of the model, through the list \verb+all_couplings+. Finally, the list \verb+all_EWOperators+  collects additional objects relevant for the calculation of electroweak Sudakov corrections.

\subsection{Counterterms} \label{subsec:counterterms}
% !TEX root = UFO_Paper.tex

This section includes brief definitions of both the $R_2$ and UV counterterms relevant for NLO UFO models, and we additionally discuss aspects relevant to the automation of the computation of loop amplitudes.

\subsubsection{$R_2$ counterterms}

In $d$ dimensions, one-loop amplitudes can be generically written as
\be
\ol{A}\left(\ol{q}\right) = \frac{1}{\left(2\pi\right)^4}\int {\rm d}^d\ol{q}\ \frac{\ol{N}\left(\ol{q}\right)}{\ol{D}_0\ol{D}_1\dots\ol{D}_{m-1}}\,,
\ee
where $\ol{D}_i\equiv\left(\ol{q}+p_i\right)^2-m_i^2$ are the propagator denominators with $m_i$ being the masses of the particles in the loop, $q$ is the loop momentum and $p_i$ are linear combinations of external momenta. The bar denotes all the quantities living in $d$ dimensions ($\ol{x}$), which can thus be split in a four-dimensional part ($x$) and a $d-4$ dimensional part ($\tilde x$) in dimensional regularisation, $\ol{x} \equiv x+\tilde x$.

Rational terms are finite contributions generated by the integration over $d-4$ pieces of the one-loop integrand. They are organised into two sets of contributions called $R_1$ and $R_2$. The rational terms $R_1$ originate from the $d-4$ component of the integrand denominators, and they can be computed similarly as the four-dimensional part of the integrand but using a different basis of scalar integrals~\cite{Ossola:2008xq}.
The $R_2$ terms are instead due to the $d-4$ component of the numerator
\be
R_2\equiv\lim_{\epsilon\to 0}\frac{1}{\left(2\pi\right)^4}\int {\rm d}^d\ol{q}\ \frac{\tilde{N}\left(\tilde{q},q,\epsilon\right)}{\ol{D}_0\ol{D}_1\dots\ol{D}_{m-1}}\label{eq:r2def},
\ee
where $d\equiv 4-2\epsilon$ and $\tilde{N}\left(\tilde{q},q,\epsilon\right)\equiv \bar{N}(\bar{q})-N(q)$. Various schemes exist for the definition of the rational terms. For example, in the 't Hooft-Veltman scheme~\cite{tHooft:1972tcz} all the quantities involved in the loop, \ie\ the loop momentum, the Dirac matrices and the metric, are taken as living in $d$ dimensions, so that 
\be\bsp
  \ol{\eta}^{\ol{\mu}\,\ol\nu}\ol\eta_{\ol\mu\,\ol\nu} = d\qquad\text{and}\qquad
  \ol{\gamma}^{\ol\mu} {\ol\gamma}_{\ol\mu}  = d\,\identity,
\esp\ee
where $\identity$ denotes the identity matrix in the Dirac space. Instead, the external momenta and the polarisation vectors live in four dimensions. Another scheme dependence is related to the choice of properties of the matrix $\gamma_5$ in $d$ dimensions.\footnote{This scheme is only relevant when considering axial anomalies.} For example, the Dirac matrices in $d$ dimensions $\ol{\gamma}_{\ol\mu}$ can be chosen to anti-commute with $\gamma_5$~\cite{Kreimer:1993bh,Korner:1991sx,Kreimer:1989ke}. In this case, the cyclic property of a Dirac trace has to be dropped to avoid algebraic inconsistency. 

An extra scheme has to be defined when computing $R_2$ terms related to operators including more than two fermions due to the presence of evanescent operators, namely operators which are only non-vanishing in $d\neq4$ dimensions. Any Lorentz invariant four-fermion operator can always be decomposed in a basis of four-fermion operator,
\be
\bar{f_1}\Gamma_a f_2 \bar{f_3}\Gamma'_a f_4=\sum_k (b_k+a_k \varepsilon) \bar{f_1}\tilde\Gamma_k f_2 \bar{f_3}  \tilde\Gamma'_k f_4\,,
\label{eq:fourfermions}\ee
where $f_i$ are fermions, $\Gamma'_a$ and $\Gamma_a$ are products of Dirac matrices that appear in the operator considered, and $\tilde\Gamma_k$ and $\tilde\Gamma'_k$ are products of Dirac matrices that define a basis of four-fermion operators in four dimensions. In the above expression, only the $b_k$ coefficients are fixed by requiring that both sides are equal in four dimensions. The determination of the coefficients $a_k$ in \eqref{eq:fourfermions} requires extra conditions to be imposed, as for instance by requiring the equality of the trace of the Dirac structure~\cite{Buras:1989xd}
\be
{\rm Tr} \Big(\tilde\Gamma_m \Gamma_a \tilde\Gamma'_m \Gamma'_a\Big) = \sum_k (b_k+a_k \varepsilon) {\rm Tr} \Big( \tilde\Gamma_m \tilde\Gamma_k \tilde\Gamma'_m \tilde\Gamma'_k\Big)\,.
\ee
Once a scheme is fixed, the integral~\eqref{eq:r2def} can be evaluated from a set of process-independent Feynman rules which can be computed once and for all in a given model. The $R_2$ terms can typically not be captured by a direct four-dimensional implementation of the numerator of all possible loop integrands, and they must therefore be computed separately and analytically.

Finally, we emphasise that the $R_2$ and $R_1$ terms are not separately gauge-invariant, but only their sum is. This provides a means for a mutual check of the implementation of the model and the package employed for NLO computations, as $R_1$ and $R_2$ terms are computed independently. The former is handled by the NLO tool whilst the latter is provided in the NLO UFO model.

\subsubsection{UV counterterms}
\label{sec:UVct}
In general, loop amplitudes in a quantum-field theory are not finite. One type of related divergences originates from loop-momenta with large Euclidean norm, and these divergences are usually referred to as {\it ultraviolet divergences}. They should be removed by the well-known renormalisation procedure, which reabsorbs them into a redefinition of the tadpoles, the fields and the free parameters of the model provided that the Lagrangian is renormalisable.\footnote{Renormalisable is understood here in a wide way, such that effective field theories are considered renormalisable, but order by order in the effective scale expansion.} This yields
\be\bsp
t^\phi_0 \to &\ t^\phi+\delta t^\phi\,,\\
\phi_0   \to &\ (1+\frac{1}{2}\delta Z_{\phi\phi})\phi + \sum_\chi \frac{1}{2} \delta Z_{\phi\chi} \chi\,,\\
x_0      \to &\ x+\delta x \label{eq:renodef}\,, 
\esp\ee
where $t^\phi$ is the tadpole for the field $\phi$, \ie\ the coefficient of the term linear in $\phi$ in the Lagrangian, $\phi$ and $\chi$ are physical fields with the same quantum numbers, and $x$ is an external parameter (internal parameters being subsequently renormalised through their dependence on the external parameters). An additional zero subscript denotes the bare quantities compared to the renormalised fields or parameters, and a $\delta$ precedes the renormalisation constants. In the above expression, the wave function renormalisation constants have been expanded at one loop, and we consider that each fermion chirality is renormalised independently, as in the general case fermionic matter is chiral. 

The bare Lagrangian is then the sum of a renormalised Lagrangian, depending only on renormalised quantities, and a counterterm Lagrangian at least linear in the renormalisation constants,
\be
\mathcal{L}_0=\mathcal{L}+\delta\mathcal{L}\,.
\ee
The UV counterterm vertices originate from the counterterm Lagrangian $\delta {\cal L}$, and they must be provided in an NLO UFO model. Their implementation is split within the files \verb+CT_vertices.py+, \verb+CT_couplings.py+ and \verb+CT_parameters.py+. As shown later, those vertices can be efficiently expressed in term of the renormalisation constants of the model thanks to the instantiation of the latter as {\tt CTParameters} objects. 

While the ultraviolet poles of the renormalisation constants are fixed by requiring the cancellation of the UV-divergences appearing in the loop amplitudes, their finite part can be chosen according to the renormalisation scheme considered. Some schemes are particularly suitable for numerical computations and make the evaluation of the loop amplitudes faster. For instance, imposing on-shell and/or complex renormalisation conditions for the derivation of the mass and wave function counterterms can avoid the computation of on-shell two-point loop Feynman diagrams on the external legs. Similarly, tadpole renormalisation allows us to ignore tadpole diagrams together with their renormalisation, as they identically cancel each other.

The renormalisation scheme used for the external parameters of the model has to be chosen adequately as well. Depending on the chosen scheme, the parameters may acquire a dependency on the renormalisation scale driven by related renormalisation group equations, and the associated running has to be included in order to guarantee formal NLO accuracy. The renormalisation of the strong coupling constant is a bit peculiar for hadronic collisions in that it must be set equal to what was used when determining the parton density functions. The other model parameters, such as for example the coefficients of higher-dimensional operators appearing in an effective field theory, can also run and mix through their renormalisation group equations, unless specific renormalisation conditions are chosen (see for example \cite{Das:2016pbk} for a fixed scale renormalisation of the new physics couplings without running). 

Finally, we mention that special counterterms such as those related to the restoration of supersymmetry that is explicitly broken by dimensional regularisation can also be included in NLO UFO models, in a similar way as what is performed for the UV countertems of the model~\cite{Degrande:2015vaa,Frixione:2019fxg}.

\subsection{The complex mass scheme} \label{subsec:complex_mass_scheme}
In order to properly treat unstable particles that appear in the $S$-matrix, a convenient scheme, the Complex Mass (CM) scheme, has been proposed, and it relies on the introduction of complex masses for all unstable particles~\cite{Denner:1999gp,Denner:2005fg}. The generic support of the Complex Mass scheme in NLO calculations requires a careful analytic continuation of all loop integrals defining the renormalisation counterterms, a topic about which an extensive discussion can be found in \cite{Frederix:2018nkq}.

\subsubsection{Complex-mass and on-shell renormalisations}

Imposing a renormalisation scheme leads to mass ($\delta M^2$) and wave function ($\delta Z$) renormalisation constants from the self-energy Feynman diagrams associated with the different particles. On-shell (OS) renormalisation conditions for a stable particle of (renormalised) mass $M$ yield
\be\bsp
  \delta M_{\rm OS}^2 = &\ \Re{[\Sigma(p^2=M^2)]}\,,\\
  \delta Z_{\rm OS}   = &\ -\Re{[\Sigma^{\prime}(p^2=M^2)]}\,,
\esp\ee
where the real part operator $\Re$ is only applied to the absorptive part of the self-energy function $\Sigma(p^2)$. In the CM scheme and for an unstable particle of (renormalised) mass $M$ and width $\Gamma$, the renormalisation conditions lead to
\be\bsp
  \delta M_{\rm CM}^2 = &\  \Sigma(p^2=M^2-i\Gamma M)\,,\\
  \delta Z_{\rm CM}   = &\ -\Sigma^{\prime}(p^2=M^2-i\Gamma M)\,.
\esp\label{eq:cmsconds}\ee
The renormalised complex mass hence becomes
\begin{equation}
M^2-i\Gamma M=M_0^2-\delta M_{\rm CM}^2\,,
\end{equation}
where $M_0$ is the bare mass.

One obvious distinction between the two schemes is the application of the operator $\Re$ in the OS scheme, that is absent in the CM scheme. This consideration suggests that a single UFO model compatible with both renormalisation schemes can be achieved, provided we introduce a new special function {\tt recms} to be defined in the file {\tt function\_library.py}. This function is defined by
\begin{verbatim}
recms = Function(
  name       = 'recms',
  arguments  = ('cms_cond','z'),
  argstype   = ('bool', 'complex'),
  expression = '(z if cms_cond else z.real)'
)
\end{verbatim}
In addition, a switch called {\tt CMSParam} is instantiated in the file {\tt parameters.py} as a new internal \verb+Parameter+ object. Its value should be changed according to whether the complex-mass scheme is turned on or off, which is achieved for {\tt CMSParam=0.0} and {\tt CMSParam=1.0} respectively.

\subsubsection{Analytic continuation}
A dynamic choice of an appropriate Riemann sheet is mandatory within complex renormalisation conditions for the particle masses and wave functions, which critically depends on the mass spectrum and decay table in a model. 

For example, the complex mass renormalisation constant associated with an unstable particle of mass $M$ and width $\Gamma$ is derived from its one-loop self-energy function $\Sigma(p^2=M^2-i\Gamma M)$, as shown in \eqref{eq:cmsconds}. Let us assume that there is a contribution to this self-energy function originating from a two-point scalar function $B_0$ (depending on a single mass $M_2$ and width $\Gamma_2$),
\be\label{eq:sigmacontr}\bsp
  & \Sigma(p^2=M^2-i\Gamma M) \supset \\
  & \qquad B_0(p^2; 0, M_2^2-i\Gamma_2 M_2)|_{p^2=M^2-i\Gamma M}\,.
\esp\ee
The analytic expressions for the $B_0$ integral in the first Riemann sheet, \ie\ when the imaginary part of the momentum squared $\Im{(p^2)}\geq 0$, read~\cite{Frederix:2018nkq}:
\be\bsp
  &\frac{1}{i\pi^2}B_0(p^2; 0,0) = 
     \frac{1}{\epsilon_{\rm UV}} + 2 - \log{\frac{-p^2-i0}{\mu^2}}\,,\\
  &\frac{1}{i\pi^2}B_0(p^2; 0,m^2) = 
     \frac{1}{\epsilon_{\rm UV}} + 2 + \log{\frac{\mu^2}{m^2}} \\
     &\qquad\quad + \frac{m^2-p^2}{p^2}\log{\frac{m^2-p^2-i0}{m^2}}\,,\\
  & \frac{1}{i\pi^2}B_0(p^2; m_1^2,m_2^2) =
     \frac{1}{\epsilon_{\rm UV}} + 2 - \log{\frac{p^2-i0}{\mu^2}}\\
     &\qquad\quad + \sum_{i=\pm}{\left[\gamma_i\log{\frac{\gamma_i-1}{\gamma_i}-\log{\left(\gamma_i-1\right)}}\right]}\,.
\esp\ee
In these expressions, we have explicitly indicated the UV origin of the divergence (through $1/\epsilon_{\rm UV}$), $\mu$ stands for the regularisation scale, and we have introduced
\be\bsp
  & \gamma_{\pm} = \frac{1}{2}\left(\gamma_0\pm \sqrt{\gamma_0^2-4\gamma_1}\right)\,,\\
  & \gamma_0= 1+\frac{m_1^2}{p^2}-\frac{m_2^2}{p^2}\,,\qquad
    \gamma_1= \frac{m_1^2-i0}{p^2}\,.
\esp\ee

In order to properly analytically continue the $B_0$ function appearing in \eqref{eq:sigmacontr}, a second Riemann sheet should be selected when the imaginary part of the momentum squared $\Im{(p^2)}<0$. This allows for a correct evaluation of the logarithm and square root functions. In our specific example, this gives
\be\bsp
 & \frac{1}{i\pi^2}B_0(M^2-i\Gamma M; 0,M_2^2-i\Gamma_2M_2) = 
    \frac{1}{\epsilon_{\rm UV}}+2\\
 &\quad +\log{\frac{\mu^2}{M_2^2\!-\!i\Gamma_2M_2}}+\frac{M_2^2\!-\!i\Gamma_2M_2\!-\!M^2\!+\!i\Gamma M}{M^2-i\Gamma M}\\
 &\quad \times \left\{\begin{array}{l}
    \log_{-1}\frac{M_2^2-i \Gamma_2 M_2-M^2+i \Gamma M}{M_2^2-i\Gamma_2 M_2} \\ 
      \qquad\qquad \text{if~} M\!>\!M_2 \text{~and~} \Gamma M_2>\Gamma_2 M\,,\\
  \log\frac{M_2^2-i\Gamma_2 M_2-M^2+i\Gamma M}{M_2^2-i\Gamma_2 M_2}
 \qquad \text{otherwise}.\\
\end{array} \right.
\esp\ee
In the above expression, when $M>M_2$ and $\Gamma M_2>\Gamma_2 M$ we need to evaluate the logarithm in the second negative Riemann sheet,
\be\bsp
\log_{-1}{z} \equiv&\ \log{z}-2\pi i\,,\\
\log{z}      \equiv&\ \log{|z|} + i \arg{z} ~~\text{with}~ -\!\pi<\arg{z}\leq \pi\,.
\esp\ee
Moreover, we mention that the analytic continuation of the general $B_0$ function (\ie\ $B_0(a; b,c)$ with $a\neq 0$, $b\neq 0$ and $c\neq 0$) is more involved than that achieved in the example. 

In order to render a UFO model compliant with the CM scheme, we first define, for practical purpose, the finite remainder $B_{0f}$ of the function $B_0$,
\begin{equation}
 B_{0f}(a; b,c) \equiv \frac{1}{i\pi^2}B_0(a; b,c)-\left(\frac{1}{\epsilon_{\rm UV}} \!+\! 2 \!+\! \log{\frac{\mu^2}{c}}\right),
\end{equation}
that is independent of the regularisation scale $\mu$. We then include, in the file {\tt function\_library.py}, the instantiation of a \verb+Function+ object \verb+B0F+ that can be used to calculate $B_{0f}(a; b,c)$ when $a\neq 0$ and $c\neq 0$.\footnote{We recall that $B_{0f}(a; b,c)$ is symmetric in $b$ and $c$, so that $B_{0f}(a; b,0)=B_{0f}(a; 0,b)$.} All other cases are simpler and could be explicitly ignored in terms of a UFO implementation, as the resulting expressions can be written in terms of $\log{}$ and $\log_{(\pm)}{}$ functions with
\begin{equation}
  \log_{(\pm)}{z}  \equiv \log{z}\pm 2\pi i \ \theta\big(-\Re{(z)}\big) \ \theta\big(\mp \Im{(z)}\big)\,.
\end{equation}
These logarithmic functions must however be included in a UFO model, and they are thus all defined in the file {\tt function\_library.py} as {\tt reglog} (for $\log{}$), {\tt reglogp} (for $\log_{(+)}{}$) and {\tt reglogm} (for $\log_{(-)}{}$). For instance, the declaration of the {\tt reglog} function reads
\begin{verbatim}
reglog = Function(
  name       = 'reglog',
  arguments  = ('z'),
  argstype   = ('complex'),
  expression = '(0.0 if z==0.0
     else cmath.log(z))'
)
\end{verbatim}

We emphasise that an alternative and more robust method to evaluate all these two-point functions could rely on the {\it trajectory method} proposed in \cite{Frederix:2018nkq}. They should then be implemented as low-level functions, directly in the directories \verb+Fortran+ or \verb+Cpp+ of the UFO. An explicit example as implemented in the \mgamc\ framework has been reported in~\cite{Pagani:2020rsg}, following the algorithm outlined in~\cite{Frederix:2018nkq}. 
\subsection{Electroweak Sudakov corrections}\label{subsec:ew}
In the present section, we briefly review the formalism relevant for the evaluation of electroweak Sudakov corrections in the leading and subleading logarithmic approximation, or in other words, in the high-energy expansion of any given observable in powers of $\log \frac{s}{m^2_W}$ where $s$ is the partonic centre-of-mass energy and  $m_W$ is the mass of the $W$ boson. A one-loop-accurate algorithmic procedure for calculations in this approximation has been available for a long time ago~\cite{Denner:2000jv}, and it has been automated in recent years for SM processes \cite{Bothmann:2020sxm, Pagani:2021vyk}. Such calculations can be  achieved via a few basic ingredients: the eigenvalues and eigenvectors of specific electroweak operators, and the finite but logarithmic-enhanced contributions to the parameter and wave-function renormalisation constants. Those quantities have been hard-coded, for example in the \mgamc\ code \cite{Pagani:2021vyk}, and their values had to be consistently matched to the conventions of the Feynman rules used in the calculations of the related amplitudes. This step can be avoided by providing the necessary additional information directly within a UFO model, using the same conventions as those used for the computation of amplitudes. Moreover, while so far the calculation of electroweak corrections has been automated for SM processes only, in the case of the Sudakov approximation it would be in principle also achievable for many BSM processes as soon as this additional information would be provided within UFO models. 

It is important to bear in mind that for the calculation of electroweak corrections in the Sudakov approximation, the UFO model does not need to contain all the information necessary for the exact evaluation of NLO electroweak corrections. Within their approximate version, electroweak corrections can be extracted from the knowledge of two kinds of ingredients, namely the finite but logarithmically-enhanced contributions to the parameter and wave-function renormalisation constants, and the eigenvalues and eigenvectors of relevant electroweak operators. The latter can be derived from the content of the model in terms of gauge eigenstates and their associated representation. However, they are rarely available for an NLO UFO model expressed in terms of mass eigenstates. Consequently, this requires some addition to the NLO UFO format, that are to be implemented in the file \verb+CT_ewcasimirs.py+. 
On the contrary, the finite but logarithmic-enhanced contributions to the parameter and wave-function renormalisation constants can be either computed once and for all from the knowledge of the bare Lagrangian and stored in the file \verb+CT_parameters.py+, already described above, or they can be identified within the complete expressions of UV counterterms usually available in an NLO UFO model allowing for exact NLO EW corrections. In this second case, it is only necessary to provide additional information allowing for the selection of the finite but logarithmic-enhanced component of the correction.\footnote{In~\cite{Bothmann:2020sxm, Pagani:2021vyk}, vertex counterterms have not been used to evaluate electroweak corrections in the Sudakov approximation originating from parameter renormalisation. Alternative numerical methods, involving {\it e.g.} the derivative of the relevant amplitudes, have been employed instead. These methods were employed in the first place because of missing information about vertex counterterms, which can now be suitably implemented in the file \texttt{CT\_vertices.py}.}

In the following we briefly describe the analytical structure of the Denner-Pozzorini (DP) algorithm \cite{Denner:2000jv}, following as much as possible the notation introduced in~\cite{Pagani:2021vyk}. Practical details are provided at the end of section~\ref{subsec:ctufo}.

We consider a process involving $n$ external particles that we identify through the indices $i_1$, $\ldots$, $i_n$ and momenta $p_1$, $\ldots$, $p_n$. The associated tree-level amplitude is denoted by ${\cal M}_0^{i_1 \ldots i_n}$, and the one-loop electroweak Sudakov corrections $\delta {\cal M}^{i_1 \ldots i_n}$ can be written as 
\begin{equation}\label{eq:onlytree}
\delta {\cal M}^{i_1 \ldots i_n}=\delta^{\rm EW}_{i'_1 i_1 \ldots i'_n i_n} {\cal M}_0^{i'_1 \ldots i'_n}\,.
\end{equation}
In this expression, the matrix elements ${\cal M}_0^{i'_1 \ldots i'_n}$ are tree-level amplitudes associated with processes with $n$ external particles that include up to two particles different from these of the original process. In addition, $\delta^{\rm EW}_{i'_1 i_1 \ldots i'_n i_n}$ collects contributions involving logarithms or double logarithms of kinematic invariants of the process, and of the squared mass of the $W$ boson ($m^2_W$) and $Z$ boson ($m^2_Z$).

The contributions to $\delta^{\rm EW}$ can be classified as
\begin{equation}\label{eq:ewgeneral}
  \delta^{\rm EW}  =  \delta^{\rm LSC}  + \delta^{\rm SSC}  + \delta^{\rm C}  + \delta^{\rm PR} \,.
\end{equation}
In this expression, we have organised the corrections in their leading ($\delta^{\rm LSC}$) and subleading ($\delta^{\rm SSC}$) soft-collinear logarithmic contributions, the purely collinear logarithmic terms ($\delta^{\rm C}$), and the logarithms originating from parameter renormalisation ($\delta^{\rm PR}$). 

The leading soft-collinear terms can be expressed as a sum over all external legs,
\begin{equation}
  \delta^{\rm LSC} {\cal M}^{i_1 \ldots i_n} = \sum_{k=1}^n \delta^{\rm LSC}_{i'_k i_k}(k) {\cal M}^{i_1 \ldots i_k' \ldots i_n}\,,
\end{equation}
where the correction factors $\delta^{\rm LSC}_{i'_k i_k}$ depend on the properties of all possible pairs of states $i_k$ and $i_k'$ that can couple via $SU(2)_L\times U(1)_Y$ interactions. These read~\cite{Kuhn:1999de, Fadin:1999bq}
\begin{equation}\begin{split}
   \delta^{\rm LSC}_{i'_k i_k} = &\ -\frac{\alpha}{8\pi} \bigg[
        \delta_{i'_k i_k} Q^2_k L^{\rm EM}
      + C^{\rm EW}_{i'_k i_k} \log^2\frac{s}{m_W^2} \\
    &\qquad\quad - 2 \big(I^Z_{i'_k i_k}\big)^2 \log\frac{m_Z^2}{m_W^2}\log\frac{s}{m_W^2}
   \bigg]\,,
\end{split}\end{equation}
where $L^{\rm EM}$ collects all logarithms of purely electromagnetic origin below the $m_W$ scale, and $Q_k$ is the electric charge of the state $i_k$. The $\delta^{\rm LSC}$ factor can thus be automatically derived, for any given process in any given model, once the eigenvalues of the effective Casimir operator matrix $C^{\rm EW}$ and these of the $I^Z$ operator matrix are known, together with information on how the different states couple via electroweak interactions.

The subleading soft-collinear contributions are given as a double sum over the pairs of external states that can couple via electroweak interactions, each term in the sum featuring the exchange of a specific neutral or charged electroweak vector boson $V=A, Z, W^\pm$. They read
\begin{equation}
  \delta^{\rm SSC} {\cal M}^{i_1 \ldots i_n} = \sum_{k=1}^n \sum_{\ell<k} \sum_V \delta^{V, \rm SSC}_{i'_k i_k i'_\ell i_\ell} {\cal M}^{i_1 \ldots i_k' \ldots i_\ell' \ldots i_n}\,,
\end{equation}
where the individual photon, $Z$ and $W$ boson corrections are respectively given by~\cite{Pagani:2021vyk}
\begin{equation}\begin{split}
  \delta^{A, \rm SSC}_{i'_k i_k i'_\ell i_\ell}  =  & \frac{\alpha}{2\pi} \log\frac{s}{Q^2} \left(\log\frac{|r_{k\ell}|}{s}-i \pi \Theta(r_{k\ell})\right) I^A_{i'_k i_k} I^A_{i'_\ell i_\ell}\,,\\
  \delta^{Z, \rm SSC}_{i'_k i_k i'_\ell i_\ell}  =  & \frac{\alpha}{2\pi} \log\frac{s}{m_W^2} \left(\log\frac{|r_{k\ell}|}{s}-i \pi \Theta(r_{k\ell})\right) I^Z_{i'_k i_k} I^Z_{i'_\ell i_\ell}\,,\\
  \delta^{W^\pm, \rm SSC}_{i'_k i_k i'_\ell i_\ell}  =  & \frac{\alpha}{2\pi} \log\frac{s}{m_W^2} \left(\log\frac{|r_{k\ell}|}{s}-i \pi \Theta(r_{k\ell})\right) I^\pm_{i'_k i_k} I^\mp_{i'_\ell i_\ell}\,.
\end{split}\end{equation}
Here, we have neglected the terms denoted as $\Delta^{s\rightarrow r_{k\ell}}$ in~\cite{Pagani:2021vyk}, since they are not relevant for the present discussion. Moreover, the scale $Q$ is the regularisation scale related to photon infrared divergences (we assume $Q^2\sim s$), $r_{kl} = (p_k + p_\ell)^2$ and $\Theta$ denotes the usual Heaviside step function. As for the leading logarithms, the correction factors are hence generically known for any process in any model as soon as the eigenvalues of the $I^A$, $I^Z$ and $I^\pm$ operator matrices are provided.

We now turn to the single logarithmic contributions arising from the soft or collinear regime. These are written as a single sum over the external legs of the process,
\begin{equation}\label{eq:deltac}
  \delta^{\rm C} {\cal M}^{i_1 \ldots i_n} = \sum_{k=1}^n \Big[ \delta^{\rm coll}_{i'_k i_k} + \frac12 \delta Z_{i'_k i_k}\Big]{\cal M}^{i_1 \ldots i_k' \ldots i_n}\,.
\end{equation}
Such corrections are derived from the wave-function renormalisation constants relevant to all external legs in the process (possibly involving mixing with fields sharing the same quantum numbers), as well as from the mass-singular loop diagram contributions that respectively read, for bosons and fermions~\cite{Denner:2001gw},
\begin{equation}\begin{split}
  \delta^{\rm B,  coll}_{i'_k i_k} =&\ \frac{\alpha}{4 \pi} C^{\rm EW}_{i'_k i_k}\log\frac{Q^2}{m_W^2}\,,\\
  \delta^{\rm F,  coll}_{i'_k i_k} =&\ \frac{\alpha}{2 \pi} \bigg[ C^{\rm EW}_{i'_k i_k}\log\frac{Q^2}{m_W^2} + Q_k^2 \log\frac{m_W^2}{m_k^2} \bigg] \,.
\end{split}\end{equation}
In these equations, $m_k$ and $Q_k$ are respectively the mass and the electric charge of the (fermionic) state $k$, while $Q$ stands for the regularisation scale. The $\delta^{\rm C}$ corrections can hence be evaluated generically for any process in any model once the eigenvalues and eigenvectors of the electroweak effective Casimir operator $C^{\rm EW}$ and the finite but logarithmic-enhanced component of the wave-functions renormalisation constants  $\delta Z_{i'_k i_k}$ are known.

The last contributions in \eqref{eq:ewgeneral} arise from the renormalisation of the input parameters of the model, like $\alpha$ (or $G_F$), $m_W$, $m_Z$ and the masses of the Higgs boson and of the top quark. These logarithmic corrections are analogous to those associated with the wave-function renormalisation constants in \eqref{eq:deltac}, and include the finite but logarithmically-enhanced component of the parameter renormalisation constants. In both cases these have to be calculated externally, and  they must be then provided as instances of the \verb+CTParameter+ and \verb+CTVertex+ classes. 
\subsection{Counterterm implementation}\label{subsec:ctufo}
% !TEX root = UFO_Paper.tex

We reviewed in section~\ref{subsec:counterterms} the origin of the UV and $R_2$ counterterms that are relevant for numerical NLO calculations, and how they should be defined in order to obtain correct results from an NLO UFO model. In this section, we discuss the details of the format in which such counterterms are explicitly specified. One particular difficulty that an OLP faces when making use of UFO counterterms is to ensure that their selection is consistent with the loop diagrams contributing to the loop amplitude considered. This task can indeed be complicated by the fact that OLP users are often given the freedom to filter out any gauge-invariant selection of loop diagrams. It is therefore desirable that the OLP is able to enforce a strict correspondence between the loop diagrams generated and the associated counterterm. 

For this reason, an NLO UFO model offers the possibility to group counterterms originating from a given subset of loop diagrams. A specific counterterm is then identified by a few properties that include the list of particles attached to the loop, the non-repeating set of particles running in the loop(s) `corresponding' to this counterterm, the cumulative coupling orders (see section~\ref{sec:interactions}) appearing in the loop vertices, and a keyword identifying the nature of the counterterm as well as how it is intended to be matched to the contributing loops.

We start our discussion of these properties with the example of the QCD $R_2$ counterterm for the triple-gluon vertex, encoded in the class {\tt CTVertex}, that therefore features a few new attributes relative to the class \verb+Vertex+ previously introduced:
\begin{verbatim}
V_R23G = CTVertex(
  name           = 'V_R23G',
  particles      = [ P.G, P.G, P.G ],
  color          = [ 'f(1,2,3)' ],
  lorentz        = [ L.VVV1 ],
  loop_particles = [ 
    [ [P.u],[P.d],[P.c],[P.s],[P.b],[P.t] ],
    [ [P.G] ] 
  ],
  couplings      = {
        (0,0,0):C.R2_3Gq, 
        (0,0,1):C.R2_3Gg
  },
  type = 'R2'
)
\end{verbatim}
The triple nested structure of the {\tt loop\_particles} attribute allows users to group together similar contributions, while retaining the correct counterterm multiplicity in the case where only a subset of particles is selected at the level of the OLP. In this example, the $R_2$ counterterms stemming from \emph{each} fermion species are all equal to the coupling {\tt R2\_3Gq} labelled by the key {\tt (0,0,0)} in the \verb+couplings+ attribute. Here, the third index refers to the position in the {\tt loop\_particles} list while the first two indices refer to the particular colour and Lorentz structure considered (as for standard vertices in a UFO model; see section~\ref{sec:interactions}).

The {\tt type} of the counterterm encodes not only its nature but also how the OLP should match it to the loop diagrams present in the computation. $R_2$ counterterms are specified by setting the attribute \verb+type+ to the value \verb+R2+. By construction and for a given set of external particles, one such counterterm must be included for each possible loop in the model, \ie\ for each loop with the specific external particles considered that has given particles running in it, and with given cumulative coupling orders. Counterterms with the attribute \verb+type+ set to the value \verb+UVmass+ take their name from the \emph{mass} renormalisation constants, and they feature a similar one-to-one correspondence to the (two-point) loop diagrams generated by the OLP. Their matching is therefore performed exactly like for counterterms of type \verb+R2+, the \verb+type+ keyword serving to encode this time the UV nature of the counterterm. On the other hand, counterterms for which \verb+type+ attribute is fixed to the value \verb+UVloop+ (or \verb+UV+) should not be considered by the matching procedure. The {\tt loop\_particles} attribute is in this case only used to discard the counterterm if any of its specified loop particles appears as having been globally excluded from the process definition by the user. Finally, counterterms for which the attribute \verb+type+ is set to \verb+UVtree+ do not have a direct correlation to any particular loop diagram, and as such they should be constructed independently. Their contribution must be built exactly like that of tree-level diagrams, while however enforcing the presence of \emph{exactly one} such counterterm vertex per diagram. This type of counterterm is for example well suited to implement counterterms restoring supersymmetry when it is explicitly broken by dimensional regularisation.

The need for a distinction between the types {\tt UVmass} and {\tt UVloop} may seem unnecessary at first, so we illustrate its use-case with the UV QCD counterterm $Z_{gd\bar{d}}$ of the \emph{vertex} $gd\bar{d}$. Such a counterterm is defined by
\begin{equation}
\label{eq:vertex_renormalisation}
    Z_{gd\bar{d}} \equiv Z_{\alpha_s}^{\frac{1}{2}} Z_{g}^{\frac12} Z_{d}^{\frac{1}{2}} Z_{\bar{d}}^\frac{1}{2}\,,
\end{equation}
which depends on the wave function renormalisation constants associated with the particles incoming to the vertex and with that of the relevant coupling. This highlights the fact that the loop particles associated with the counterterm $Z_{gd\bar{d}}$ are not directly related to the corresponding loop diagrams, but rather to the loop particles originating from the explicit calculation of the involved renormalisation constants. This leads to a lack of direct correspondence between a vertex counterterm and its corresponding loop corrections. The listed {\tt loop\_particles} cannot thus be \emph{matched} directly to the particle content of the contributing diagrams, and they only represent an overall list of allowed particles in the process. According to the UFO standard, this counterterm is defined with type {\tt UV} $\equiv$ {\tt UVloop},
\begin{verbatim}
V_UVGDD = CTVertex(
  name           = 'V_UVGDD',
  particles      = [ P.d__tilde__,P.d,P.G ],
  color          = [ 'T(3,2,1)' ],
  lorentz        = [ L.FFV1 ],
  loop_particles = [  
        [ [P.u],[P.d],[P.s] ],
        [ [P.c] ],
        [ [P.b] ],
        [ [P.t] ],
        [ [P.G] ]
    ],
  couplings      = {
      (0,0,0):C.UV_GQQq,(0,0,1):C.UV_GQQc,
      (0,0,2):C.UV_GQQb,(0,0,3):C.UV_GQQt,
      (0,0,4):C.UV_GQQg
  },
  type = 'UV'
)
\end{verbatim}
A more detailed and technical description of the \emph{vertex} renormalisation and loop particle matching procedure can be found in eqs.~(2.80)--(2.87) in \cite{Alwall:2014hca}.

We now discuss the implementation of the couplings assigned to {\tt CTVertex} instances. The main difference with respect to tree-level couplings already used in LO-only UFO is the necessity of supplying them in the form of terms of a Laurent series in the dimensional regulator $\epsilon$. This is achieved by allowing for the specification of an \emph{expansion dictionary} as the {\tt value} of the coupling,\footnote{All entries not specified in the dictionary representation of the Laurent series are to be understood as being zero. Moreover, there is no restriction on the range of $\epsilon$ orders spanned by the Laurent series.}
\begin{equation}
    \frac{A}{\epsilon^2} + \frac{B}{\epsilon} + C \leadsto {\tt \{-2:A, -1:B, 0:C\}}
\end{equation}
Given the omnipresence of wave function renormalisation constants in all vertex counterterms (see \eqref{eq:vertex_renormalisation}), it is desirable to be able to define counterterm-related \emph{parameters} which are themselves expansions in $\epsilon$. This is allowed in the NLO UFO format, and taken advantage of, for example when writing the coupling {\tt UV\_GQQt} appearing in the definition of the \verb+CTVertex UV_GQQt+,
\begin{verbatim}
UV_GQQt = Coupling(
  name  = 'UV_GQQt',
  value = 'complex(0,1)*G_UVt*G',
  order = {'QCD':3}
)

G_UVt = CTParameter(
  name  = 'G_UVt',
  type  = 'real',
  value = {
    -1 : '((G**2)/(96.0*cmath.pi**2))*4.0*TF',
     0 : 'cond(MT, 0.0,
            -((G**2)/(96.0*cmath.pi**2))
               *4.0*TF*reglog(MT**2/MU_R**2))'
  },
  texname = '\delta Gt'
)
\end{verbatim}
where the new {\tt cond} function, implemented in the file {\tt function\_library.py}, is a shortcut function designed to support the cases of both zero and finite top mass. When the first argument of the \verb+cond+ function is equal to 0, then the second argument of the \verb+cond+ function is returned. Otherwise, the third argument is returned. Moreover, the coupling order in this example is {\tt QCD=3}, which corresponds to the cumulative coupling orders of the loop corrections to that vertex. It is important that this coupling order is correctly set, as it may be used in the OLP matching procedure when building the counterterm contributions.

The introduction of the {\tt CTParameter} object named {\tt G\_UVt} is convenient (though not necessary), as it will appear in the counterterms associated with many QCD vertices. In order to facilitate the usage and import of the model, there are a few limitations to writing counterterm couplings. First, the {\tt value} attribute of a counterterm coupling can be either a string or an expansion dictionary. When the latter is used, the string expression of its values can only involve instances of {\tt Parameter}, but not of {\tt CTParameter}. Second, when writing the coupling value as a string, it must correspond to a term whose summands each contain \emph{at most} $k$ occurrences of {\tt CTparameter} objects for a UFO model suitable for N$^{k}$LO computations. At NLO, the expression \verb"'2*ParamA*CTParamB + 4*CTParamE'" would be acceptable (one \verb+CTParameter+ instance in each of the two summands), but \verb"'2*CTParamA*CTParamB + 4*CTParamE'" would not (two \verb+CTParameter+ instances in the first term).

Additionally, the UFO 2.0 format does not explicitly differentiate the UV and infrared quantities $\epsilon_{\rm UV}$ and $\epsilon_{\rm IR}$ in the expansion dictionaries. The distinction between them can however be retained at NLO by using reserved parameters named {\tt epsUV} and {\tt epsIR} as multiplicative factors, that are defined as external parameters in a Les Houches block \verb+TECHNICAL+.\footnote{The UV-finite part of the wave function counterterm of massless fermions typically includes poles in the IR regulator $\epsilon_{\rm IR}$.} In addition, NLO UFO models include a \verb+LOOP+ block that is reserved for the renormalisation scale parameter \verb+MU_R+ that appear in loop integrals.

Before closing this section, we now discuss how to implement in an NLO UFO model the information necessary for the calculation of electroweak Sudakov corrections to any matrix element in the leading and subleading logarithmic approximation. As detailed in section~\ref{subsec:ew}, these corrections can be generically derived for any process from the knowledge of the eigenvalues of the electroweak effective Casimir operator matrix $C^{\rm EW}$, that of the photon, $Z$-boson and $W$-boson operator $I^A$, $I^Z$ and $I^\pm$, and from information on which pairs of states interact through electroweak interactions (and of course how).  Additional quantities such as the coefficients of the related beta function or Dynkin operators may also be useful in this context, and can be further added to this list (see \cite{Denner:2000jv}).

In the following, we first consider as an example the matrix associated with the electroweak Casimir operator when it acts on neutral weak bosons, ~\cite{Denner:2000jv}
\begin{equation}
  C^{\rm EW}_{AZ} = \frac{2}{s_w^2} 
    \begin{pmatrix} s_w^2 & -s_wc_w\\ -s_wc_w & c_w^2\end{pmatrix}\,.
\end{equation}
In this expression, $s_w$ and $c_w$ stand for the sine and cosine of the electroweak mixing angle. The corresponding implementation in a UFO model is achieved through the declaration of an \verb+EWOperator+ object,
\begin{verbatim}
CEW_AZ = EWOperator(
  name      = 'CEW_AZ',
  type      = 'casimir',
  particles = [ [P.A, P.Z] ],
  elements    = {
    (0,0): P.EW_AA, (0,1): P.EW_AZ, 
    (1,0): P.EW_ZA, (1,1): P.EW_ZZ 
  }
)
\end{verbatim}
In this example, the \verb+EWOperator+ object considered is defined through four mandatory attributes. The first of them consists of its name (\verb+name+), and the second of them, \verb+type+, can either take the value \verb+'casimir'+ (for the eigenvalues of the effective electroweak Casimir operator, like in the above example) or refer to one of the electroweak bosons of the theory (for the various $I^V$ operators, as illustrated in the next example below). The value of the attribute \verb+particles+ is a list that provides information on the states relevant for the operator considered. This primary list includes a single list for the electroweak Casimir operator $C^{\rm EW}$ and the `neutral' operators $I^A$ and $I^Z$, and two lists for the `charged' operators $I^\pm$ that connect different elements of the weak multiplets of the model. The (non-zero) matrix elements of the operator are finally provided through the value of the attribute \verb+elements+, that contains a dictionary mapping any non-zero element of the matrix to a \verb+Parameter+ object (declared as detailed in section~\ref{sec:parameters}).

As another example, we provide a possible implementation of the $I^+$ operator, for the case in which it acts on left-handed quarks,\footnote{In \cite{Denner:2000jv}, the CKM matrix is approximated to a unit matrix, so that only the diagonal elements of the quark $I^\pm$ matrices are non-zero. In our example, we consider the general case.}
\begin{verbatim}
Ip = EWOperator(
  name      = 'Ip',
  type      = P.W__plus__,
  particles = [ 
    [ P.u, P.c, P.t], [P.d, P.s, P.b]
  ],
  chirality = 'left',
  elements  = {
    (0,0):C.EW_ud,(0,1):C.EW_us,(0,2):C.EW_ub, 
    (1,0):C.EW_cd,(1,1):C.EW_cs,(1,2):C.EW_cb, 
    (2,0):C.EW_td,(2,1):C.EW_ts,(2,2):C.EW_tb
  }
)
\end{verbatim}
This time the attribute \verb+particles+ contains two lists of particles, as the $I^+$ operator relates up-type fermions (the first list, \verb+[ P.u, P.c, P.t]+) and down-type ones (the second list, \verb+[ P.d, P.s, P.b]+). Moreover, an optional attribute (\verb+chirality+), relevant when fermions are involved, has been included in the declaration. Its value  indicates that only the left-handed components of the fermions involved is concerned.

In addition to the information provided above, the derivation of all logarithmic corrections to any given matrix element shown in \eqref{eq:ewgeneral} requires to include those stemming from parameter and wave-function renormalisation. This can be achieved automatically once the renormalisation constants are computed or, in the case of a model that allows for electroweak corrections in the Sudakov approximation but not for exact NLO electroweak corrections, they have to be calculated for this purpose. In both cases, it is necessary to access directly the finite but logarithmically-enhanced component of the parameter counterterms and possibly their impact in vertex counterterms.  The analytical results should therefore be provided, through declarations of \verb+CTParameter+ and possibly \verb+CTVertex+ objects relevant to the implementation of counterterms (see above).

\section{Conclusion}\label{sec:conclusions}
In this paper, we have presented the current 2.0 update of the UFO format for (B)SM models, that we have coined the `\textit{Universal Feynman Output}' format. This new name has been adopted to distinguish the current format from its initial version, as the UFO has evolved significantly during the last decade. Moreover, the UFO is not solely connected with \feynrules\ anymore, but it lies at the heart of many high-energy physics software tools. 

The UFO 2.0 format includes several new features that were not part of the initial proposal, thanks to the flexible and modular structure that drove the design of the UFO ten years ago, and that allows it to be easily expandable and encompass features relevant for the interest of high-energy physics software at a given time. Initially, the UFO format has been designed to include information on a model's particles, the list and values of the parameters appearing in the model's Lagrangian and the associated interaction vertices.  In addition to such information, UFO 2.0 models can optionally include information on the particle's decay widths, on the renormalisation group running of the model's parameters and masses, and on ingredients relevant for automatic higher-order perturbative calculations. Moreover, users can include form factors and enforce the usage of custom propagators in their implementation. Whereas some of these features were already described in earlier publications, others have never been documented officially in any scientific article. 

It was the aim of the present paper to release the most up-to-date documentation of the UFO format, collecting in a single document information about all features that could be present in a UFO model, from the initial mandatory ones to those subsequently developed during the last decade.

\section*{Acknowledgments}
This work has been partly supported by grants from the French ANR (grants ANR-21-CE31-0013 `DMwithLLPatLHC' and ANR-20-CE31-0015 `PrecisOnium'), the Deutsche Forschungsgemeinschaft (grant 499573813 `EFTTools', grant 396021762--TRR 257 and Germany's Excellence Strategy-EXC 2121 `Quantum Universe' -- 3908333), the German Federal Ministry for Education and Research (BMBF contract 05H21WWCAA), the ERC (grant 101041109 `BOSON'), the CNRS IEA (grant 205210 `GlueGraph'), the European Union's Horizon 2020 research and innovation program (grant 824093 STRONG-2020, EU Virtual Access `NLOAccess'), the F.N.R.S (MAXLHC IISN convention 4.4503.16), and the Fermi National Accelerator Laboratory (Fermilab), a U.S.\ Department of Energy, Office of Science, HEP User Facility managed by Fermi Research Alliance, LLC (FRA), and acting under Contracts No.\ DE--AC02--07CH11259.

\bibliography{UFOBib}

\end{document}